\documentclass[a4paper,12pt]{article}
\usepackage{ifpdf}
\usepackage{a4wide}
\usepackage{amsfonts}
\usepackage{mathrsfs}
\usepackage{amssymb}
\usepackage{amsmath}
\usepackage{color}
\usepackage[
      colorlinks=true,
      linkcolor=blue,
      urlcolor=blue,
      filecolor=black,
      citecolor=blue,
            ]{hyperref}
            
\usepackage{amsfonts}
\usepackage{graphicx}
\usepackage{epstopdf}
\usepackage{bigstrut,bigdelim,multirow}

\begin{document}
\title{Coexistence and competition of ferromagnetism and p-wave superconductivity in holographic model}
\author{\large
~Rong-Gen Cai~\footnote{E-mail: cairg@itp.ac.cn}~,
~~Run-Qiu Yang~\footnote{E-mail: aqiu@itp.ac.cn}
\\
\small State Key Laboratory of Theoretical Physics,\\
\small Institute of Theoretical Physics, Chinese Academy of Sciences,\\
\small Beijing 100190,  China  }

 \maketitle
\begin{abstract}
By  combining  a holographic p-wave superconductor model and a holographic  ferromagnetism model, we study the coexistence and competition of ferromagnetism and p-wave superconductivity. It is found that the results depend on the self-interaction of magnetic moment of the complex vector field and which phase appears first. In the case that the ferromagnetic phase appears first, if the interaction is attractive, the system  shows the ferromagnetism and superconductivity can coexist in low temperatures.  If the interaction is repulsive,  the system will only be in a pure ferromagnetic state.  In the case that the superconducting phase appears first, the attractive interaction will leads to a magnetic p-wave superconducting phase in low temperatures. If the interaction is repulsive, the system will be in a pure p-wave superconducting phase or ferromagnetic phase when the temperature is lowered.
\end{abstract}
\newpage
\tableofcontents

\section{Introduction}
In condensed matter physics, there are two kinds of critical phenomena which have attracted a lot of attention for a long time.  One is ferromagnetism where the electron spins align to produce a net magnetization, which breaks the time reversal symmetry spontaneously and happens in the ferromagnets at the Curie temperature, $T_C$ (sometimes it is even higher than the indoor temperature). The other  is  superconductivity where electrons condense into Cooper pairs in the momentum space, which breaks the U(1) symmetry spontaneously and usually happens at a much low temperature $T_{sc}$. Though ferromagnetism has been found for more than thousands years and the superconductivity was found  about 100 years ago, their physical explanations was provided after the complete quantum theory on materials was built.

In a very long time, it was thought these two phenomenons are incompatible with each other. This is rooted in the microscopic theory of superconductivity  by Bardeen, Cooper, and Schrieffer (BCS)~\cite{Bardeen}. In the framework of BCS theory, the superconductivity  appears when the electrons were bounded with antiparallel spins in singlet Cooper pairs due to the effective attractive force coming from the lattice vibrations.
When magnetic impurity atoms are placed in a conventional superconductor, they are capable of flipping the electron's spin. Hence, impurity will suppresses the singlet Cooper pair formation, which causes a rapid depression of the superconducting transition temperature $T_{sc}$. Likewise, the superconductivity can screen off the magnetic field, which leads to the long-range magnetic order is accompanied by the expulsion of superconductivity. These features lead to the competition of such two orders. However, around 1980s, it was recognized that under special conditions superconductivity may coexist with antiferromagnetic order, where neighboring electron spins arrange in an antiparallel configuration. For instance, in heavy fermion antiferromagnets, the itinerant magnetic moments have almost no de-pairing effect on singlet Cooper pairs, because the average exchange interaction is zero.

The discovery of the first superconducting ferromagnet\footnote{In this paper, we will use ``superconducting ferromagnet" to denote the materials whose Curie temperature is higher than superconducting transition temperature and ``ferromagnetic superconductor"  to denote the opposite case.} UGe$_2$ in the year 2000 came as a big surprise~\cite{Lonzarich}. In this material, superconductivity is realized well below the Curie temperature, without expelling the ferromagnetic order. Since then, other two  superconducting ferromagnets have been discovered, such as URhGe~\cite{Aoki} and UCoGe~\cite{Huy}, which  display intrinsic coexistence of ferromagnetism and superconductivity. Evidence of superconducting ferromagnetic phase was also reported for ZrZn$_2$ in 2001~\cite{Pfleiderer}.

\begin{figure}
\includegraphics[width=1.1\textwidth]{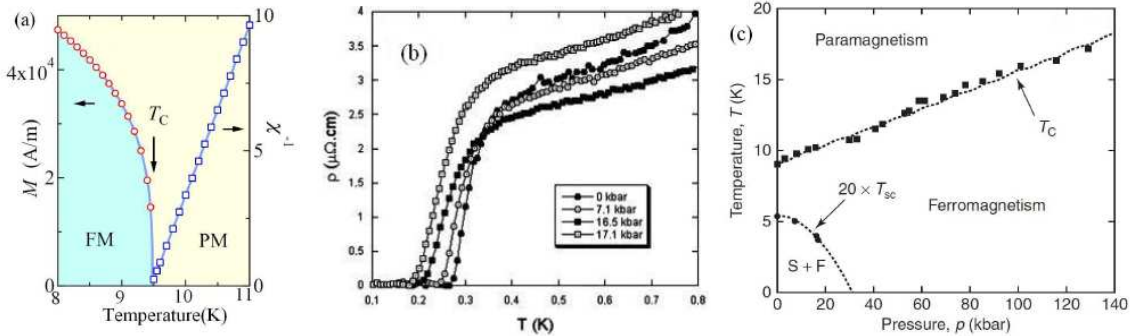}
\caption{The experimental results on the p-wave superconducting ferromagnetic material URhGe. {\bf (a)} Temperature dependence of the spontaneous magnetization $M(T)$ and the inverse of the  magnetic susceptibility $\chi^{-1}(T)$ under the normal pressure~\cite{Naoyuki}. {\bf (b})Temperature dependence of the resistivity of URhGe for different pressures at low temperature. The superconducting critical temperature is defined by the zero resistivity point~\cite{F.Hardy}. {\bf (c)} Pressure-temperature phase diagram of URhGe~\cite{F.Hardy}.}
\label{URhGe}
\end{figure}

Figure~\ref{URhGe} shows the experimental results of a typical superconducting ferromagnetic material URhGe. The ferromagnetic order is observed near $T_C\simeq9.5K$, where the spontaneous magnetization begins appearing and the magnetic susceptibility diverges. Below this temperature, the initial magnetic susceptibility is positive and the spontaneous magnetization increases with decreasing the temperature. In the ferromagnetic phase, when temperature is decreased to near $T_{sc}\simeq0.25K$(at atmospheric pressure), the resistivity approaches  to  zero, which shows that the material is in a superconducting phase. In the region of $T<T_{sc}$, spontaneous magnetization and zero resistivity coexist, which shows a superconducting ferromagnetism. The $P-T$ phase diagram shows that this kind coexistence can appear in a very wide pressure region (up to near 30 kbar, where the $T_{sc}$ disappears.).

The nature of superconducting state in ferromagnetic materials is currently under debate. Early investigations~\cite{Karchev} studied the coexistence of conventional s-wave superconductivity with itinerant ferromagnetism. However, the scenario of spin-triplet pairing soon gained the upper hand~\cite{MacHida}. For a review of phenomenological theory of ferromagnetic unconventional superconductors with spin-triplet Cooper pairing of electrons, one can see  Ref.~\cite{D.I.Uzunov}.  Ref.\cite{US} presents a general thermodynamic theory that describes phases and phase transitions of ferromagnetic superconductors with spin-triplet electron Cooper pairing, based on an extended Ginzburg-Landau theory. Generally speaking, the coexistence of ferromagnetism and superconducting is discovered in spin triplet rather than the usual spin singlet superconducting materials,  because that p-wave pairing allows parallel spin orientation of the fermion Cooper pairs in unconventional superconductors~\cite{Vollhardt:1990}. This unconventional paired manner makes p-wave superconductivity being robust under the influences of external magnetic field and spontaneous magnetization. So they may coexist with each other.

A mean-field model for coexistence of spin-triplet pairing and ferromagnetism was developed in~\cite{MacHida,Nevidomskyy}. The model considers a uniform coexistence of ferromagnetism and superconductivity, i.e., the same electrons play the role for the ferromagnetism and superconductivity at the same time. Another scenario where there is an interplay between magnetic and superconducting orders in the same material is superconductors with spiral or helical magnetic order. Examples of such include ErRh$_4$B$_4$ and HoMo$_6$S$_8$. In these cases, the superconducting and magnetic order parameters entwine each other in a spatially modulated pattern, which allows for their mutual coexistence, although it is no longer uniform. Even spin-singlet pairing may coexist with ferromagnetism in this manner.

Up to now, the theoretical investigation have been concentrated on the weak coupling case, where some approximations and  conception of free field are still valid. The investigations on strong correlated system in theoretical and experimental aspects have challenged the pictures about the materials. A crucial feature of these systems is the nonzero magnetic moment of the spin-triplet Cooper pairs. However, the microscopic theory about the coexistence of magnetism and superconductivity in strongly interacting heavy electrons is either too complex or insufficiently developed to describe the complicated behavior. So it is still a fascinating thing to find a suitable theory to describe the coexistence and competition of the ferromagnetism and superconductivity in  strong correlated system, such as the heavy fermion system~\cite{Coleman}, Iron-based superconductor~\cite{Kordyuk} and unconventional superconductor~\cite{Norman}.

In the strong coupling case, the usual methods developed in condensed matter theory (CMT) are considered losing their efficacy. A new method,  named gauge/gravity duality or AdS/CFT correspondence, is considered as a promising approach~\cite{Maldacena:1997re,Gubser:1998bc,Witten:1998qj,Witten:1998qj2}. This method relates a weak coupling gravitational theory in  a $(d+1)$-dimensional asymptotically anti de-Sitter (AdS) space-time to a $d$-dimensional strong coupling conformal field theory (CFT) in the AdS boundary. In recent years, this duality has been extensively applied into condensed matter systems. Some remarkable  progresses have been made in this direction. For example, some gravitational dual models of superfluid/superconductor~\cite{Gubser,Hartnoll:2008vx,Domenech:2010nf}, (non-)Fermi liquid~\cite{Lee:2008xf,Liu:2009dm,Cubrovic:2009ye}, Josephson junctions~\cite{Horowitz:2011dz,Kiritsis:2011zq,Wang}, superconducting quantum interference device~\cite{Cai:2013sua} and magnetic properties in superconductors~\cite{Montull:2009,Donos:2012yu,Albash:2008eh,Montull:2012fy,Iqbal:2010eh} have been constructed and intensively studied. The models in the AdS/CFT frame for ferromagnetism/paramagnetism and anti-ferromagnetism/paramagnetism phase transitions have also been proposed in~\cite{Cai:2014oca,Cai:2014jta}.

In this paper, we will explore the coexistence and competition between ferromagnetism and p-wave superconductivity in the strong coupling case. Our tool is just the gauge/gravity duality. We noted that the similar topic  appeared in Ref.~\cite{Amoretti:2013oia}, where the authors use two U(1) fields condense simultaneously from SU(2) model to present the superconducting and spontaneous magnetic orders.
Our model is different from theirs.


In order to  construct a holographic model to investigate the coexistence and competition between ferromagnetism and superconductivity, we need a model where ferromagnetic properties (such as spontaneous magnetic moment, time reversal symmetry broken and diverged magnetic susceptibility at Curie temperature and so on) can appear independently from the superconductivity. This is just supplied by the model in Ref.~\cite{Cai:2014oca}. By combining the complex vector field model for the holographic p-wave superconductor~\cite{Cai:2013pda,Cai:2013aca,Cai:2014ija} and the real antisymmetric tensor field model for the holographic ferromagnetism in Ref.~\cite{Cai:2014oca}, we study the coexistence and competition of ferromagnetism and superconductivity. It turns out that the results depend on magnetic moment self-interaction of complex vector field and which phase appears first. In the case that the ferromagnetic phase appears first and the interaction is attractive, the p-wave superconductivity can still appear and the system can show ferromagnetism and superconductivity both when temperature is lower than a critical value. But if the  interaction is repulsive, the p-wave superconductivity can not appear and the system will only be in a pure ferromagnetic state.  In the case that the superconducting phase appears first, the system will show a magnetic p-wave superconducting phase with decreasing temperature if the  interaction is attractive. If the interaction is repulsive, the system is in  a pure p-wave superconducting phase or ferromagnetic phase when the temperature is lowered.

The paper is organized as follows. In section~\ref{model}, we will first describe the holographic model. We will give our ansatz for matter fields, the equations of motion (EoMs)  and the expression of free energy density in  section~\ref{freeN}. In  section~\ref{coexist}, we will investigate the possible phases and the coexistence and competition between ferromagnetism and superconductivity with different parameters. A brief summary and discussion will be given in  section~\ref{sum}.

\section{The model}
\label{model}
In this paper, the model we are considering  is the combination of the holographic p-wave superconductor model described by a complex vector field~\cite{Cai:2013pda,Cai:2013aca,Cai:2014ija} and holographic ferromagnetism model described by a real antisymmetric tensor field~\cite{Cai:2014oca}. The action is
\begin{equation}\label{action1}
S=\frac1{2\kappa^2}\int d^4x\sqrt{-g}\left[R+\frac{6}{L^2}-F_{\mu\nu} F^{\mu \nu}+\mathcal{L}_{\rho}+\mathcal{L}_{M}+\mathcal{L}_{\rho M}\right],
\end{equation}
with
\begin{equation}\label{LrhoM}
\begin{split}
\mathcal{L}_\rho&=-\frac{1}{2}\rho_{\mu\nu}^\dagger\rho^{\mu\nu}-m_1^2\rho_\mu^\dagger\rho^\mu+iq\gamma \rho_\mu\rho_\nu^\dagger F^{\mu\nu}-V_\rho,\\
\mathcal{L}_M&=-\frac14\nabla^\mu M^{\nu\tau}\nabla_\mu M_{\nu\tau}-\frac{m_2^2}4M^{\mu\nu}M_{\mu\nu}-\frac{\lambda}2M^{\mu\nu}F_{\mu\nu}-V_M,\\
\mathcal{L}_{\rho M}&=-i\alpha\rho_\mu\rho_\nu^\dagger M^{\mu\nu},\\
\end{split}
\end{equation}
where $L$ is the AdS radius which will be set to be unity and $\kappa^2\equiv 8\pi G $ is related to the gravitational constant in the bulk. In the following, we will set $2\kappa^2=1$ for simplicity. $\lambda, \gamma$ and $J$~are three constants with~$J<0~\cite{Cai:2014oca}, \gamma>0$. $m_1, m_2$ are the masses of the complex vector field  $\rho_\mu$ and real tensor field $M_{\mu\nu}$. $g$ is the determinant of the bulk metric $g_{\mu\nu}$ and $q$ is the charge of complex vector field. We define $F_{\mu\nu}=\nabla_\mu A_\nu-\nabla_\nu A_\mu$ and $\rho_{\mu\nu}=D_\mu\rho_\nu-D_\nu\rho_\mu$ with the covariant derivative $D_\mu=\nabla_\mu-iqA_\mu$. The $\gamma$'s term characterizes the magnetic moment of the vector field $\rho_\mu$~\cite{Cai:2013pda,Cai:2013kaa}. The antisymmetric tensor~$M_{\mu\nu}$~is the effective polarization tensor of the U(1) gauge field strength~$F_{\mu\nu}$. $V_M$ describes the self-interaction of the polarization tensor. Following  Ref.~\cite{Cai:2014oca}, we take
\begin{equation}\label{VM1}
V_M=\frac J8{M_\mu}^\nu {M_\nu}^\tau {M_\tau}^\delta {M_\delta}^\mu
\end{equation}
for simplicity. The term $V_\rho$ describe the self-interaction of  magnetic moment of the complex vector field, a simple form is
\begin{equation}\label{Vrho}
V_\rho=-\frac{\Theta}2 \rho_{[\mu}\rho_{\nu]}^\dagger\rho^\mu\rho^{\dagger\nu}.
\end{equation}
Here $\Theta$ is a constant, it characterizes the feature of magnetic moment interaction of complex vector field. From the free energy density given later, we can see that a positive $\Theta$ gives an attractive interaction between the magnetic moment of complex vector field itself, while a negative value of $\Theta$ gives an repulsive interaction. In  Refs.~\cite{Cai:2013pda,Cai:2013aca,Cai:2014ija,Cai:2013kaa}, the term $V_\rho$ is not  considered as there the  aim is to study the superconductivity of the model. However, we can see that this term will play an important role in the  following when we consider the magnetic properties of the model.

One may find that the Lagrangian $\mathcal{L}_M$ has a little difference from  the form in~\cite{Cai:2014oca}. This is just for convenience. In order to simplify our discussion in the probe limit, we make following transformations
\begin{equation}\label{transf2}
M_{\mu\nu}\rightarrow\lambda M_{\mu\nu},~J\rightarrow\lambda^{-2}J,~\rho_\mu\rightarrow\lambda\rho_\mu,\alpha\rightarrow\alpha/\lambda,\Theta\rightarrow\Theta\lambda^{-2}.
\end{equation}
Under these transformations,  action~\eqref{action1} can be rewritten as
\begin{equation}\label{action2}
S=\int d^4x\sqrt{-g}\left[R+\frac{6}{L^2}-F_{\mu\nu} F^{\mu \nu}+\lambda^2(\mathcal{L}_{\rho}+\mathcal{L}_{M}+\mathcal{L}_{\rho M})\right],
\end{equation}
with
\begin{equation}\label{LrhoM2}
\mathcal{L}_M=-\frac14\nabla^\mu M^{\nu\tau}\nabla_\mu M_{\nu\tau}-\frac{m_2^2}4M^{\mu\nu}M_{\mu\nu}-\frac12M^{\mu\nu}F_{\mu\nu}-V_M
\end{equation}
and the others are kept the same as the forms in~\eqref{LrhoM}. The probe limit corresponds to keeping all the quantities finite with $\lambda\rightarrow0$. In this limit, we can fix the background geometry and Maxwell field and only consider the dynamics of complex vector field and polarization field.

\section{EoMs and free energy density}
\label{freeN}
\subsection{Ansatz and EoMs}
As in Ref.~\cite{Cai:2014oca}, we will work in the probe limit with $\lambda\rightarrow0$, by which we can fix the geometry background and neglect the back reaction of matter fields on the background
geometric and Maxwell field. The equations for complex vector field and polarization field read
\begin{equation}\label{eqPM}
\begin{split}
D^\nu\rho_{\nu\mu}-m_1^2\rho_\mu+iq\gamma\rho^\nu F_{\nu\mu}+\Theta \rho_{[\nu}\rho_{\mu]}^\dagger\rho^\nu-i\alpha\rho^\nu M_{\nu\mu}=0.\\
 \nabla^2M_{\mu\nu}-m_2^2M_{\mu\nu}-J{M_\mu}^\delta {M_\delta}^\tau {M_\tau}_\nu-2i\alpha\rho_{[\mu}\rho_{\nu]}^\dagger-F_{\mu\nu}=0.
\end{split}
\end{equation}
We take the AdS  Reissner-Nordstr\"om  (RN) black hole with a planar horizon as the background metric~\cite{CaiZhang}
\begin{equation}\label{geom2}
\begin{split}
  ds^2=r^2(-f(r)dt^2+dx^2+dy^2)+\frac{dr^2}{r^2f(r)},\\
   f(r)=1-\frac{1+\mu^2}{r^3}+\frac{\mu^2}{r^4},~~~A_\mu=\mu(1-1/r)dt.
\end{split}
\end{equation}
Here the horizon radius has been set to $r_h=1$ and $\mu$ can be identified with the chemical potential in the dual field theory. The temperature of the boundary theory is
\begin{equation}\label{temp2}
T=\frac1{4\pi}(3-\mu^2).
\end{equation}
A self-consistent ansatz for complex vector field and polarization field is
\begin{equation}\label{VMansatz}
M_{\mu\nu}=-p(r)dt\wedge dr+h(r)dx\wedge dy,~~\rho_\mu=\rho_xdx+ie^{i\theta(r)}\rho_ydy.
\end{equation}
As in Ref.~\cite{Cai:2013pda}, we can take  $\rho_x(r)$ to be real by some suitable U(1) gauge. However, the $y$-component of $\rho_\mu$ then can not  be set to be real. So we have to assume that $\rho_y$ and $\theta$ are real functions depending on $r$. Taking the ansatz into equations~\eqref{eqPM}, we find that only when $e^{i\theta(r)}=\pm1$ for $\forall r\in(r_n,\infty)$ can we find a self-consistent solution. Without loss generality, we can assume $\theta(r)=0$. Considering the symmetry between $\rho_x$ and $\rho_y$,  we assume that
\begin{equation}\label{crho}
\rho_y=c(r)\rho_x
\end{equation}
in what follows. Thus we can reach the following equations for the components of matter fields
\begin{equation}\label{eqcomp1}
\begin{split}
h''+\frac{f'}fh'+\left(\frac{Jh^2}{r^6f}-\frac{2f'}{rf}-\frac4{r^2}-\frac{m_2^2}{fr^2}\right)h-\frac{2c\alpha \rho_x^2}{r^2f}=0,\\
\rho_x''+(\frac{f'}f+\frac2r)\rho_x'+\left(\frac{q^2\phi^2}{r^4f^2}-\frac{\Theta c^2\rho_x^2}{r^4f}-\frac{m_1^2}{fr^2}-\frac{ch\alpha}{fr^4}\right)\rho_x=0,\\
c''+\left(\frac{f'}f+\frac2r+\frac{2\rho_x'}{\rho_x}\right)c'-\frac{(1-c^2)(c\Theta\rho_x^2+\alpha h)}{fr^4}=0.\\
\end{split}
\end{equation}
The equation for $p(r)$ decouples from the above equations, so we will not  write down it here.


Near the AdS boundary, the linearized equations give following asymptotic solutions~\footnote{The asymptotic solution of $c(r)$ depends on the source free condition of $\rho_x$. When $\rho_{x+}\neq0$, asymptotic solution of $c(r)$ is $c=c_++c_-r^{-\delta_1}$.}
\begin{equation}\label{adsbound}
\begin{split}
\rho_x={\rho_x}_+r^{(\delta_1-1)/2}+{\rho_x}_-r^{-(\delta_1+1)/2},\quad c=c_+r^{\delta_1}+c_-,\\
h(r)=h_+r^{(1+\delta_2)/2}+h_-r^{(1-\delta)/2},
\end{split}
\end{equation}
where $\delta_1=\sqrt{1+4m_1^2}$ and $\delta_2=\sqrt{17+4m_2^2}$. According to the AdS/CFT dictionary, $\rho_{x+}$, $c_+$ and $h_+$ are sources terms for corresponding operators, while
$\rho_{x-}$, $c_-$ and $h_-$ are vacuum expectation values, respectively.  As in Ref.~\cite{Cai:2014oca}, we need impose the condition $h_+=0$ for the polarization field and ${\rho_x}_+=c_+=0$ for the complex vector field. This is consistent with the spirit of AdS/CFT correspondence: one requires that the condensation and magnetization happen spontaneously.

At the horizon, we impose regular conditions for $\rho_x,~c(r)$ and $h$, which give following relationships for the initial values at $r=r_h$
\begin{equation}\label{init1}
\begin{split}
\rho_x'=\frac1{4\pi T}\rho_x(m_1^2+c\alpha h-c\Theta\rho_x^2),\\
h'=2h-\frac{h(Jh^2-m_2^2)+2c\alpha\rho_x^2}{4\pi T},\\
c'=\frac1{4\pi T}(1-c^2)(c\Theta \rho_x^2+\alpha h).
\end{split}
\end{equation}
 Note that there exists a  symmetry as $\{\rho_x\rightarrow\rho_y,~\rho_y\rightarrow\rho_x\}$, by which we can set that $|\rho_y(r_h)|\leq|\rho_x(r_h)|$, i.e.,
\begin{equation}\label{initc}
-1\leq c(r_h)\leq1.
\end{equation}
Thus once given the value of parameters $\{\Theta,~\alpha,~J,~q,~m_1^2,~m_2^2,~T\}$ and some suitable initial values $\{h(r_h),~\rho_x(r_h),~c(r_h)\}$, we can integrate  equations~\eqref{eqcomp1} to obtain the whole solutions matching the boundary conditions $h_+={\rho_x}_+=c_+=0$ at the AdS boundary.

Our numerical results show that~\footnote{We use the function ode45 in MATLAB R2012b with the relative and absolute errors $10^{-13}$ to solve equations~\eqref{eqcomp1} numerically.}, except for case that $c(r_h)=h(r_h)=0$ or $c(r_h)=\pm1$, the integration will meet a divergency at somewhere in $r_h<r<\infty$ and the equations~\eqref{eqcomp1} do not have physical solutions. When $c(r_h)=h(r_h)=0$, according to the relationship of initial values in~\eqref{init1}, we can find that $c(r)=h(r)=0$,  which is just the p-wave superconductor solution found in  Ref.~\cite{Cai:2013pda}. When $c(r_h)=\pm1$, the solution for $c$ is $c(r)=\pm1$, i.e., $\rho_x=\pm\rho_y$, which leads that the nontrivial solutions are either pure ferromagnetic phase or p-wave superconductivity phase with nonzero magnetization if $\alpha\neq0$. From the equations~\eqref{eqcomp1}, we can see that the transformation $\alpha\rightarrow-\alpha$ is equivalent to fix $\alpha$ but make the transformation of $c(r)\rightarrow-c(r)$. So the case of $\alpha<0$ is equivalent to the case of $\alpha>0$ with exchanging the results of $c(r)=\pm1$. In the special case that $\alpha=0$, from the equations~\eqref{eqcomp1}, we can see that the equations for $h$ and $\rho_x$ are decoupled with each other. The ferromagnetic phase and p-wave superconducting phase have no interaction with each other~\footnote{This is the consequence of the probe limit. Once the back reaction is taken into account, even when $\alpha=0$, $\rho_{\mu}$ and $M_{\mu\nu}$ will interact each other through gravity background.}. We are here not interested in that case. In the following sections, therefore we will set $\alpha>0$.

\subsection{Free energy  and magnetic moment}

Let us now compute the on-shell action and give the thermodynamic of the dual boundary theory. Here, we use grand canonical ensemble by fixing the boundary chemical potential. For convenience, we can set $\mu=1$ in the numerics. The results for other chemical potential values can be obtained by scaling relations.  We will first fix the horizon radius $r_h$ to solve the equations~\eqref{eqcomp1} and compute the free energy density, then we use the scaling transformation to obtain the results in the grand canonical ensemble. In gauge/gravity duality, the Gibbs free energy $F$ can be obtained by temperature timing the on-shell bulk action with Euclidean signature. Since we work in the probe approximation, we can ignore the gravity part. Given that the system is stationary, the Euclidean action is related to the Minkowskian one by a total minus. Using the equations of motion~\eqref{eqPM} and the source free conditions, we can get the free energy density contributed by vector field and polarization field as
\begin{equation}\label{onshell2}
\frac{F}V=\lambda^2\int_{r_h}^\infty dr\sqrt{-g}\left(-V_\rho+\frac14M^{\mu\nu}F_{\mu\nu}-V_M -\frac{i\alpha}2M_{\mu\nu}\rho^\mu\rho^{\nu\dagger}\right),
\end{equation}
where $V$ is the area spanned by coordinates $x$ and $y$ on the boundary.
Taking the ansatz~\eqref{VMansatz} into the free energy density,  it turns out to be
\begin{equation}\label{onshell3}
\frac{F}V=\lambda^2\int_{r_h}^\infty dr\left(\frac{Jh^4}{4r^6}-\frac{c^2\Theta \rho_x^4}{r^2}-\frac{c\alpha\rho_x^2h}{r^2}\right).
\end{equation}
Note that the contribution to the free energy density  from $p(r)$ is not relevant to our discussions, therefore here we have neglected that part  in \eqref{onshell3}. The integration is finite when $\rho_{x+}=h_+=0$ and $m_1^2>-1/4,~m_2^2>-4$. In addition, note that  there are two symmetries in the free energy and EoMs~\eqref{eqcomp1} such as
\begin{equation}\label{symm1}
\{\rho_x\rightarrow-\rho_x,~\rho_y\rightarrow-\rho_y\},~\{c\rightarrow-c,~h\rightarrow-h\}
\end{equation}
which make we can specify  $\rho_x(r_h)\geq0$ and $h(r_h)\leq0$.  It is a general requirement that  the functions of $\rho_x$ and $h$ do not  have zero points in the region of $[r_h,\infty)$. Thus we  can assume $\rho_x(r)>0$ and $h(r) \leq0$ in the region $[r_h,\infty)$. According to the definition of magnetic moment  density and using the expression~\eqref{onshell2}, we have
\begin{equation}\label{TolM}
N=-\frac{1}V\lim_{B\rightarrow0}\left(\frac{\partial F}{\partial B}\right)_{\rho_\mu,p}=-\int_{r_h}^\infty\frac h{2r^2}dr,
\end{equation}
which is the same as in Ref.~\cite{Cai:2014oca}. By  the dictionary of AdS/CFT, the expectation value of p-wave superconducting order parameter is a complex vector $\overrightarrow{P}$, whose mode is $P=\sqrt{1+c^2}|\rho_{x-}|$.  Here it is worthwhile to mention that  though the complex vector field does not appear in the final expression of magnetic moment density, it makes contribution through the mixture terms in equations~\eqref{eqcomp1}: once $\rho_x$ and $c$ do not vanish, $h$ will not vanish.

In the pure p-wave model, when nontrivial solutions of $\rho_\mu$ appear, the global U(1) and spatial rotation symmetries are broken spontaneously. In this model, it is also true. In addition, there is an another possible symmetry breaking in this model. If one notes the fact that external magnetic field $B$ will be transformed in $-B$ under the time reversal transformation, by the expression of free energy density in~\eqref{onshell2}, we have following rules for time reversal transformation,
\begin{equation}\label{timerev}
h\rightarrow-h,\ \ \  \rho_y\rightarrow-\rho_y.
\end{equation}
So when $h\neq0$ or $\rho_y=\pm\rho_x\neq0$ (they both lead to nonzero magnetic moment), the time reversal symmetry is broken spontaneously. This agrees with the fact that a spontaneously magnetized phase will break time reversal symmetry spontaneously.

From Refs.~\cite{Cai:2013pda,Cai:2014oca}, we can see that the complex vector field and polarization field can both condense in low temperatures in an AdS RN black hole background. We take $T_{sc0}$ and $T_{C0}$ as the critical temperatures of $\rho_x$ and $h$, when $\alpha=0$,  in the AdS  RN black hole background, respectively. With decreasing the temperature of the system, three interesting questions appear immediately:

(1) If $T_{C0}>T_{sc0}$, then the system will enter into the ferromagnetic phase first. Can a p-wave superconducting phase still appear at lower temperature? If yes, is the critical temperature still $T_{sc0}$?

(2) If $T_{sc0}>T_{C0}$, then the system will enter into the p-wave superconducting phase first. Can a ferromagnetic phase appear at some lower temperature? If yes, is the critical temperature still $T_{C0}$?

(3) In an enough low temperature, for example, near the zero temperature limit, can the p-wave superconductivity and ferromagnetism coexist in this model?

In what follows, we will consider all these questions. We will first solve equations~\eqref{eqcomp1} with the regular condition~\eqref{init1} at the horizon and source free conditions at the AdS boundary. In general, depending on parameters, the system has four kinds of solutions: one is a trivial solution without vector and tensor hairs, one is a solution with vector hair, but no tensor hair (which describes the pure p-wave superconductivity phase), one is the solution with tensor hair, but no vector hair (which describes the pure ferromagnetic phase), and final one is the solution with both hairs (which describes  both the superconductivity and ferromagnetism coexistence  phase). The phase is physical favored if it has the lowest free energy.

\section{Coexistence and competition}
\label{coexist}
\subsection{Superconducting ferromagnet}

Let us  first consider the case with $T_{C0}>T_{sc0}$, i.e., the ferromagnetic phase appears first. According to the equation for $c$ in equations~\eqref{eqcomp1}, we find $c\neq0$ when $h\neq0$. So there is not a phase such that $\{h<0,~\rho_x\neq0,~\rho_y=0\}$. When we decrease the temperature to be lower than $T_{C0}$, five kinds of solutions may appear. They are phase A $\{h=\rho_x=\rho_y=0\}$, phase B $\{h<0,~\rho_x=\rho_y=0\}$, phase C $\{h=\rho_y=0,~\rho_x\neq0\}$, phase $D_1$ $\{h<0,~\rho_x=\rho_y\neq0\}$ and phase $D_2$ $\{h<0,~\rho_x=-\rho_y\neq0\}$, which corresponds to normal phase, pure ferromagnetic phase, pure p-wave superconducting phase and two kinds of superconducting ferromagnetic phases, respectively.

\begin{figure}
\includegraphics[width=0.5\textwidth]{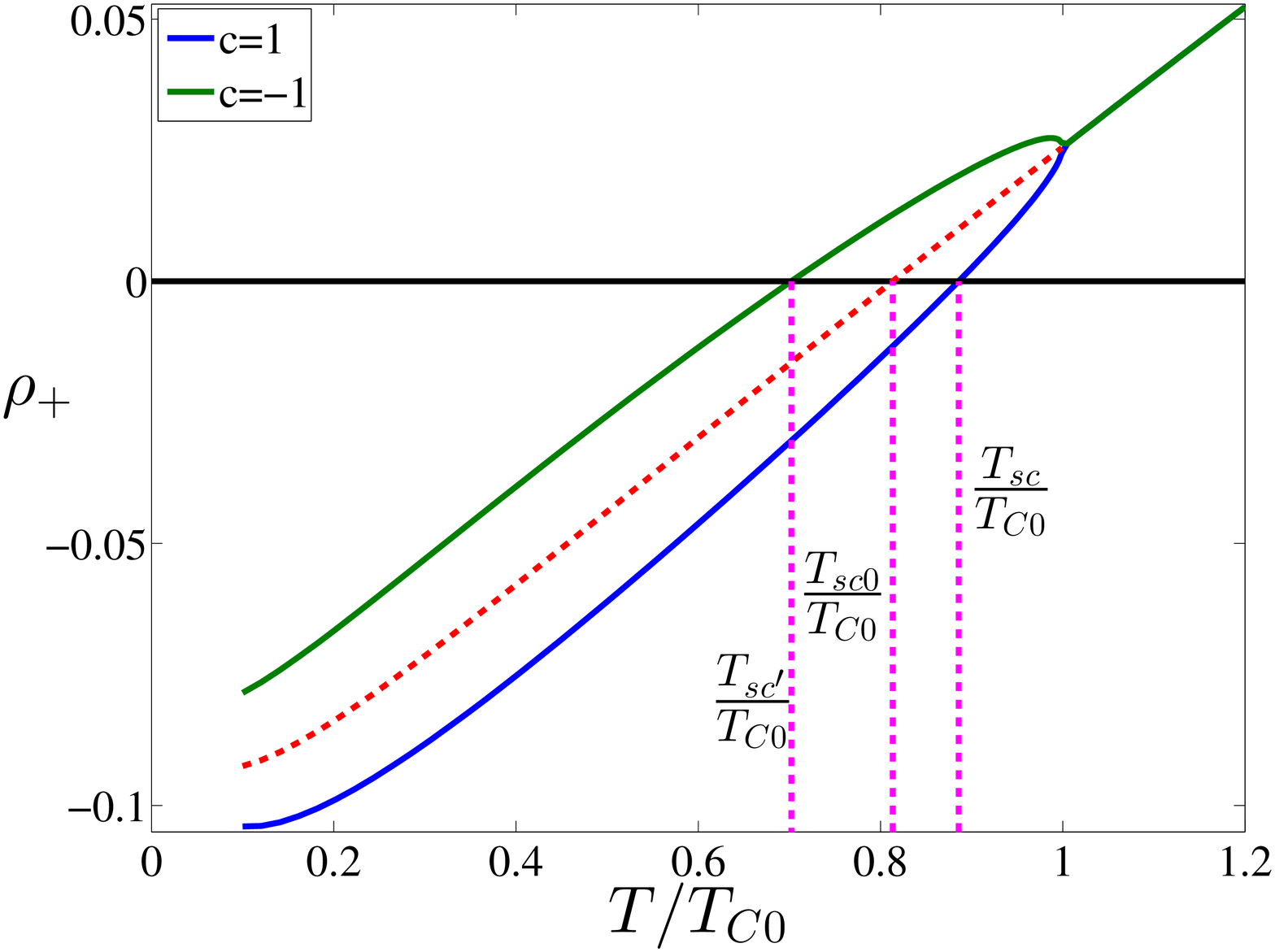}
\includegraphics[width=0.5\textwidth]{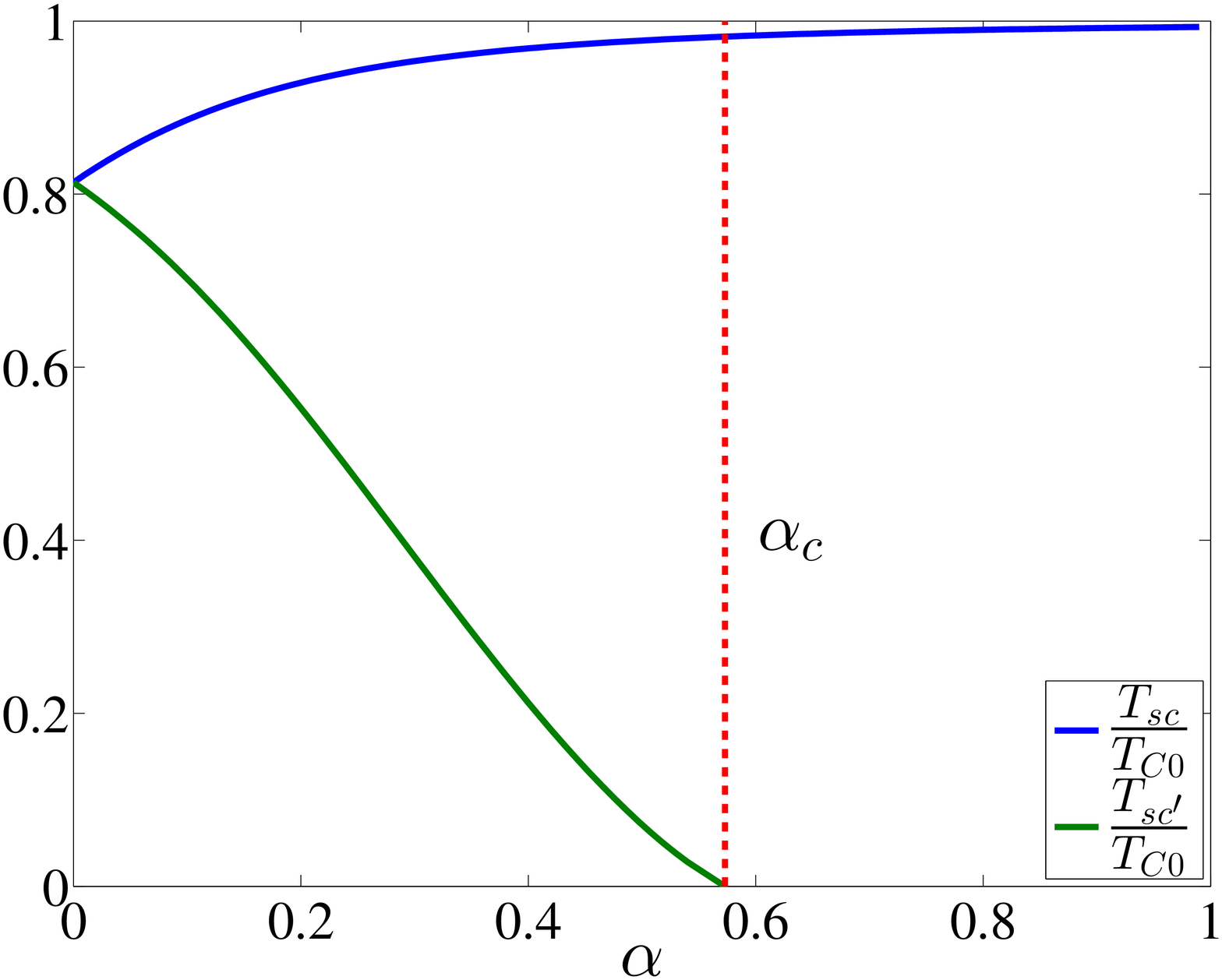}
\caption{{\bf Left:} $\rho_+$ as a function of temperature $T$ when $h<0$. The green line and blue line stand for $c=-1$ and $c=1$, respectively. The red dashed line stands for the case in the pure AdS RN background with $h=0$. {\bf Right:}  $T_{sc}/T_{C0}$ and $T_{sc'}/T_{C0}$ as  functions of $\alpha$. We take $m_1^2=-3/16, m_2^2=-3,J=-1, q=1.3$. Here $T_{C0}\simeq0.00925{\mu},~T_{sc0}\simeq0.8135T_{C0}$ }
\label{Trhoalpha}
\end{figure}

For small $\rho_x$, we can get its effective mass square from equations~\eqref{eqcomp1} as
\begin{equation}\label{effM1}
m^2_{1\text{eff}}=m_1^2-\frac{q^2\phi^2}{r^2f}+\frac{c\alpha h}{r^2}.
\end{equation}
In general, $h(r)$ does not  have zero point in the region of $r_h<r<\infty$ in the condensed phase. With the choice $h\leq0,\alpha>0$, we have $\alpha h\leq0$. If $c=1$, the effective mass square of complex vector field will decrease by the condensate of $h$. If $c=-1$, instead the effective mass square of complex vector field will  increase by the condensate of $h$, which leads  the critical temperature of complex vector field to decrease. In this case, if $-h\alpha$ is enough large, the complex vector field can not condense even in zero temperature.  These imply that when we decrease the temperature, the instability of complex vector field will appear in the manner of $\rho_y=\rho_x$ or $\rho_y=-\rho_x$ for small $\alpha$ and in the manner of $\rho_y=\rho_x$ for enough large $\alpha$.  This instability tells us that in the ferromagnetic phase, the marginally stable mode for complex vector field can still appear at some temperature less than $T_{C0}$.  The analysis can also be confirmed directly by solving equations \eqref{eqcomp1} numerically under the case of neglecting the $\rho_x^2$ terms, i.e.,
\begin{equation}\label{eqcomp2}
\begin{split}
h''+\frac{f'}fh'+\left(\frac{Jh^2}{r^6f}-\frac{2f'}{rf}-\frac4{r^2}-\frac{m_2^2}{fr^2}\right)h=0,\\
\rho_x''+(\frac{f'}f+\frac2r)\rho_x'+\left(\frac{q^2\phi^2}{r^4f^2}-\frac{m_1^2}{fr^2}-\frac{ch\alpha}{fr^4}\right)\rho_x=0,\\
\end{split}
\end{equation}
with the initial value of $\rho_x(r_h)=1$. As a typical example, we choose parameters as $m_1^2=-3/16, m_2^2=-3,J=-1$ and $q=1.3$ (for other parameter values, the results
are qualitatively similar) . The results are shown in figure~\ref{Trhoalpha}. In the left plot, we show  $\rho_+$ as a function of temperature $T$ for $c=\pm1$ in the case of $\alpha=0.1$ when $h<0$. The green line and blue line stand for $c=-1$ and $c=1$, respectively. The red dashed line stands for the case in the pure AdS  RN background with $h=0$.  In the case of $c=1$, there is a critical temperature $T_{sc}$ less than $T_{C0}$  but higher than $T_{sc0}$ to make $\rho_+=0$. In the case of $c=-1$, there is a critical temperature $T_{sc'}$ less than $T_{sc0}$ to make $\rho_+=0$. In the right plot, we show  $T_{sc}/T_{C0}$ and $T_{sc'}/T_{C0}$ as functions of $\alpha$. When we increase $\alpha$,  $T_{sc}$ will increase but  $T_{sc'}$ decreases. Numerical results show that there is a critical value at $\alpha=\alpha_{c}\simeq0.57649$, less than which $T_{sc'}$ will be less than zero. In  appendix~\ref{app1}, we  give the method to compute the critical value $\alpha_c$.

Note that equations~\eqref{eqcomp2} and figure~\ref{Trhoalpha}  only show that there is a marginally stable mode for the complex vector field at $T=T_{sc}$ or $T=T_{sc'}$. When temperature is below these critical temperatures, whether the complex vector field can condense and which one of two phases $\{h<0,\rho_x=-\rho_y\neq0\}$ and $\{h<0,\rho_x=\rho_y\neq0\}$ can appear in the physical phase space are determined by their free energy density. In order to find the phase diagram, we have to solve equations~\eqref{eqcomp1} numerically to compute the free energy of possible solutions. It turns out that the results depend on the sign of $\Theta$. The possible phases and the physical favored phase in different temperature regions are summarized in table~\ref{Tab1}.
\begin{table}
  \centering
  \begin{tabular}{|c|c|c|c|c|}
  \multicolumn{5}{c}{Phases in the case of $\alpha<\alpha_c$}\\
    \hline
    Temperature & $T>T_{C0}$& $T_{sc}<T<T_{C0}$&$T_{sc'}<T<T_{sc}$&$T<T_{sc'}$\\
    \hline
   \multirow{2}*{Possible}&  \multirow{2}*{A} &  \multirow{2}*{A,B} & A, $D_1$&A, B, $D_1$\\
   &&& $C$(if $T<T_{sc0}$) &$D_2$, C\\
    \hline
    Physical($\Theta>0$) &A& $B$ &
    \multicolumn{2}{c|}{$D_1$}\\
    \hline
    Physical($\Theta<0$) &A &
    \multicolumn{3}{c|}{B}\\
    \hline
    \multicolumn{5}{c}{ }\\
  \end{tabular}
  \\

\centering
\begin{tabular}{|c|c|c|c|}
  \multicolumn{4}{c}{Phases in the case of $\alpha>\alpha_c$}\\
    \hline
    temperature & $T>T_{C0}$& $T_{sc}<T<T_{C0}$&$T<T_{sc}$\\
    \hline
    Possible& A & A,B & A, B, $D_1$, C(if $T<T_{sc0}$)\\
    \hline
    Physical($\Theta>0$) & A& B& $D_1$\\
    \hline
    Physical($\Theta<0$) & A&
    \multicolumn{2}{c|}{B}\\
    \hline
  \end{tabular}
  \caption{The possible and physical phases in the case of $T_{C0}>T_{sc0}$. Phase A is $\{h=\rho_x=\rho_y=0\}$. Phase B is $\{h<0,~\rho_x=\rho_y=0\}$. Phase C is $\{h=\rho_y=0,~\rho_x\neq0\}$. Phase $D_1$ is $\{h<0,~\rho_x=\rho_y\neq0\}$. Phase $D_2$ is $\{h<0,~\rho_x=-\rho_y\neq0\}$. }\label{Tab1}
\end{table}

In the case of $\Theta>0$, we find that the the phase $D_1$ is physical favored when $T<T_{sc}$.  This means  that the system displays ferromagnetism and superconductivity both in low temperatures. In addition, by computing the ferromagnetic order parameters and p-wave superconducting order parameter, we see they are both continuous at two critical temperatures. With lowering  the temperature, the system will first transit into the ferromagnetic phase at $T_{C0}$ and then  into the superconducting ferromagnetic phase at $T_{sc}$ through  two second order phase transitions.  The critical temperature of complex vector field  grows  when $\alpha$ gets increased. This shows that, in the ferromagnetic state, the interaction between p-wave pairs and spontaneous magnetic moment will promote the appearance of p-wave superconductivity. On the other hand,  in the case of $\Theta<0$, though the solutions of $\rho_x\neq0$ exist, they are not physical favored because they have higher free energy than the solution of $\rho_x=0$.  As a result, in this case, there is only a ferromagnetic phase (phase B) when $T<T_{C0}$.

\begin{figure}
\includegraphics[width=0.5\textwidth]{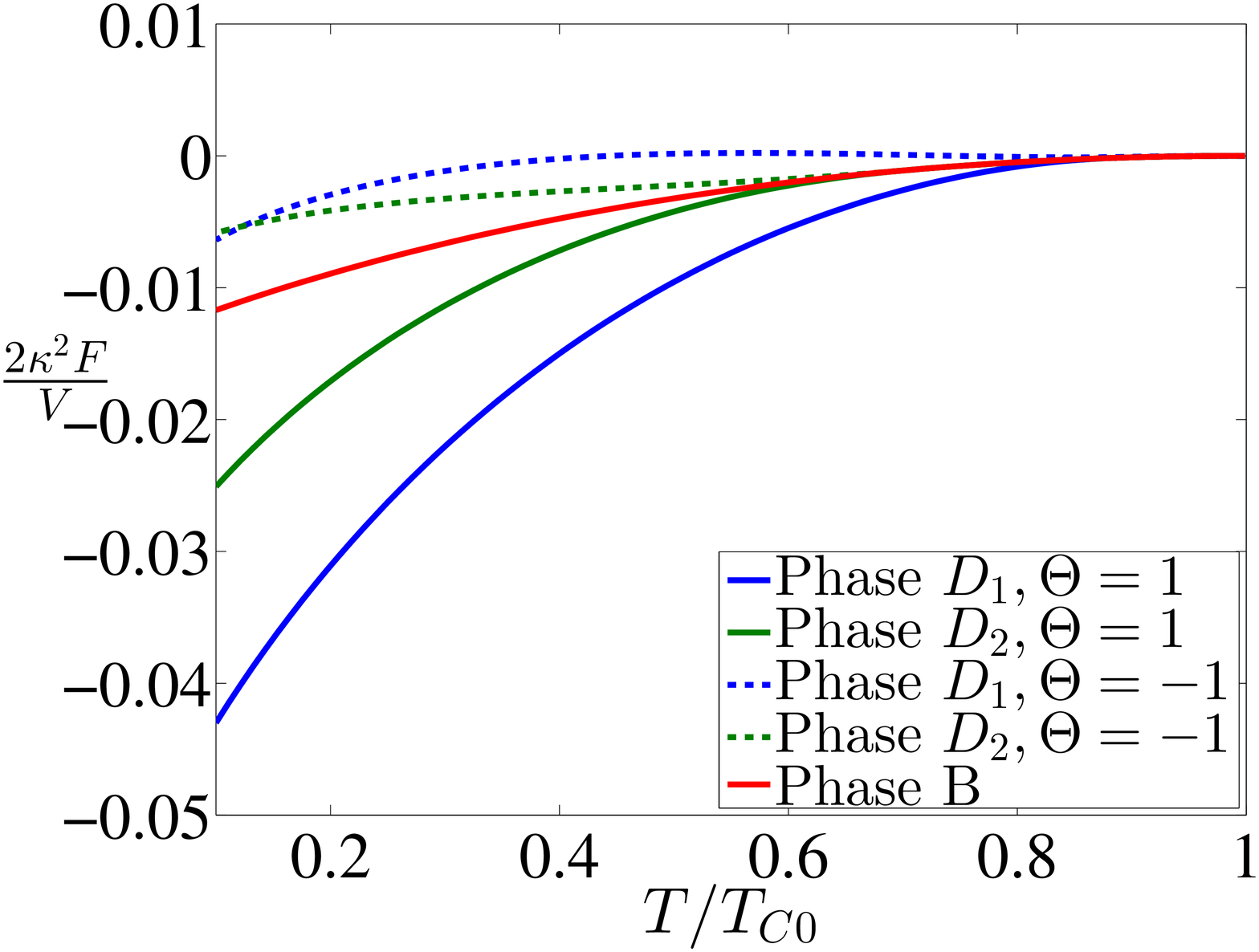}
\includegraphics[width=0.5\textwidth]{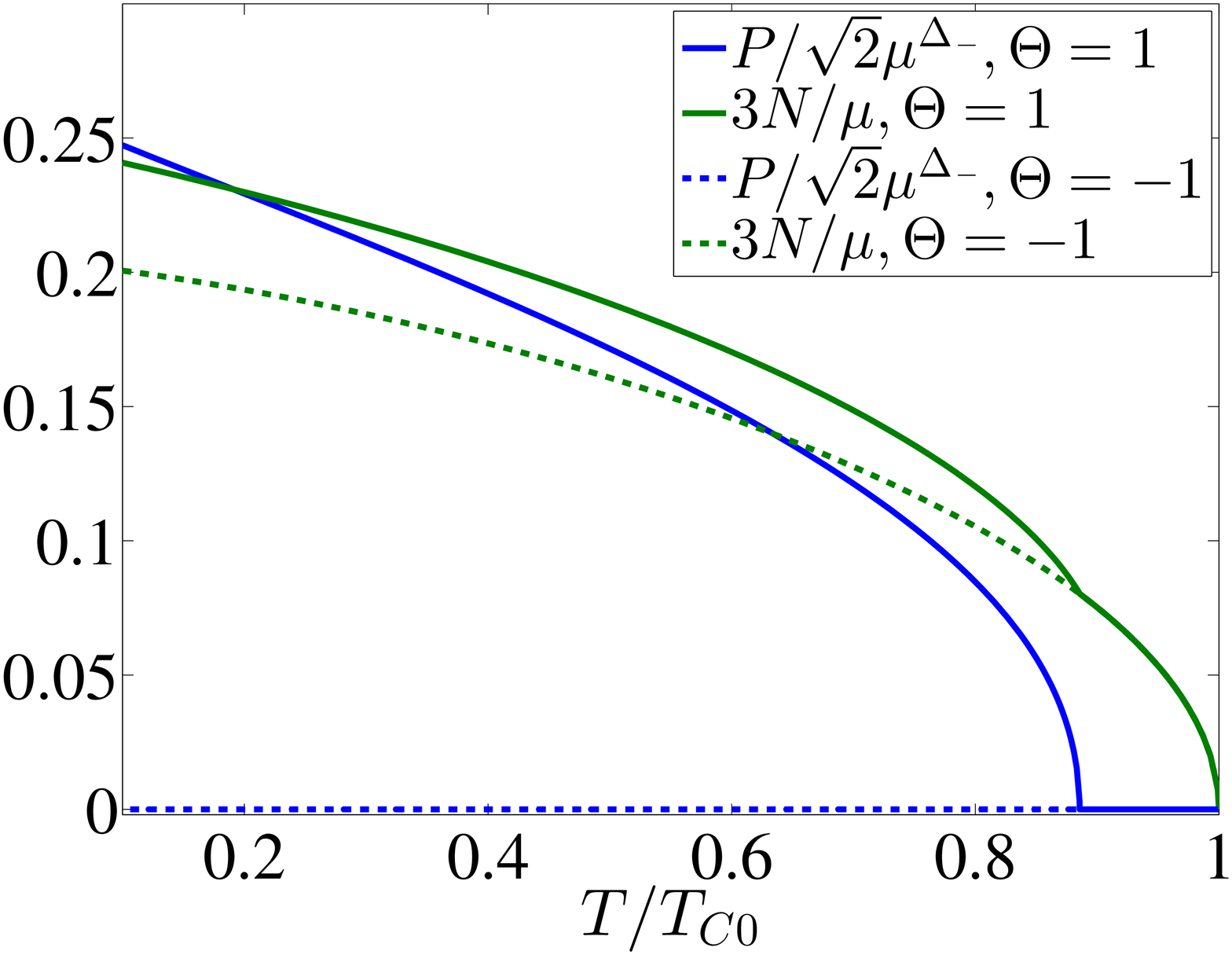}
\caption{{\bf Left:} The free energy density with respect to temperature in the case of $\Theta=\pm1$. {\bf Right:} The p-wave superconducting order parameter $P$ and spontaneous magnetic moment density $N$ as  functions of temperature in the case of $\Theta=\pm1$ in the physical favored phase. Here $m_1^2=-3/16, m_2^2=-3,J=-1, q=1.3$ and $\alpha=0.1$. $\Delta_-=1+(1+\delta_1)/2$. }
\label{TG1}
\end{figure}

Here we only show the example for the case of $\alpha<\alpha_c$. The case of $\alpha>\alpha_c$ is similar except for the difference that the solution for the phase $D_2$ does not  occur.  In the left plot of figure~\ref{TG1}, we show the free energy density of phases $D_1$ and $D_2$ in the case of $\Theta=\pm1$ and phase $B$ with respect to temperature. Note that here the on-shell free energy density of phases $A$  and $C$ is  zero in the probe approximation. One can see that phase $D_1$ and phase $B$ are physical favored when $\Theta=\pm1$ respectively when $T<T_{sc}$.  Thus we  see that the physical favored phase depends on the sign of $\Theta$. In the case of $\Theta>0$, with decreasing the temperature, the system will first goes into  the ferromagnetic phase when $T<T_{C0}$ and then into the p-wave superconducting ferromagnetic phase when $T<T_{sc}$. However,  there are two superconducting ferromagnetic phases $D_1$($\{\rho_x=\rho_y\}$) and $D_2$($\{\rho_x=-\rho_y\}$).  Which  has a lower free energy than the pure ferromagnetic phase when $T<T_{sc}$? It turns out that the physical favored one is phase $D_1$.  The numerical results show that the ferromagnetism and superconductivity can coexist in the whole region of $T<T_{sc}$. It seemingly indicates that the ferromagnetism and superconductivity can coexist even in the zero temperature limit.  However, in the case of $\Theta<0$, we  see that although the superconducting ferromagnetic phases exist (phase $D_1$ and $D_2$),  these two phases have higher free energy density than the pure ferromagnetic phase.  In this case, therefore the physical favored phase is the pure ferromagnetic phase and the p-wave superconductivity will not occur.
\begin{figure}
\includegraphics[width=0.5\textwidth]{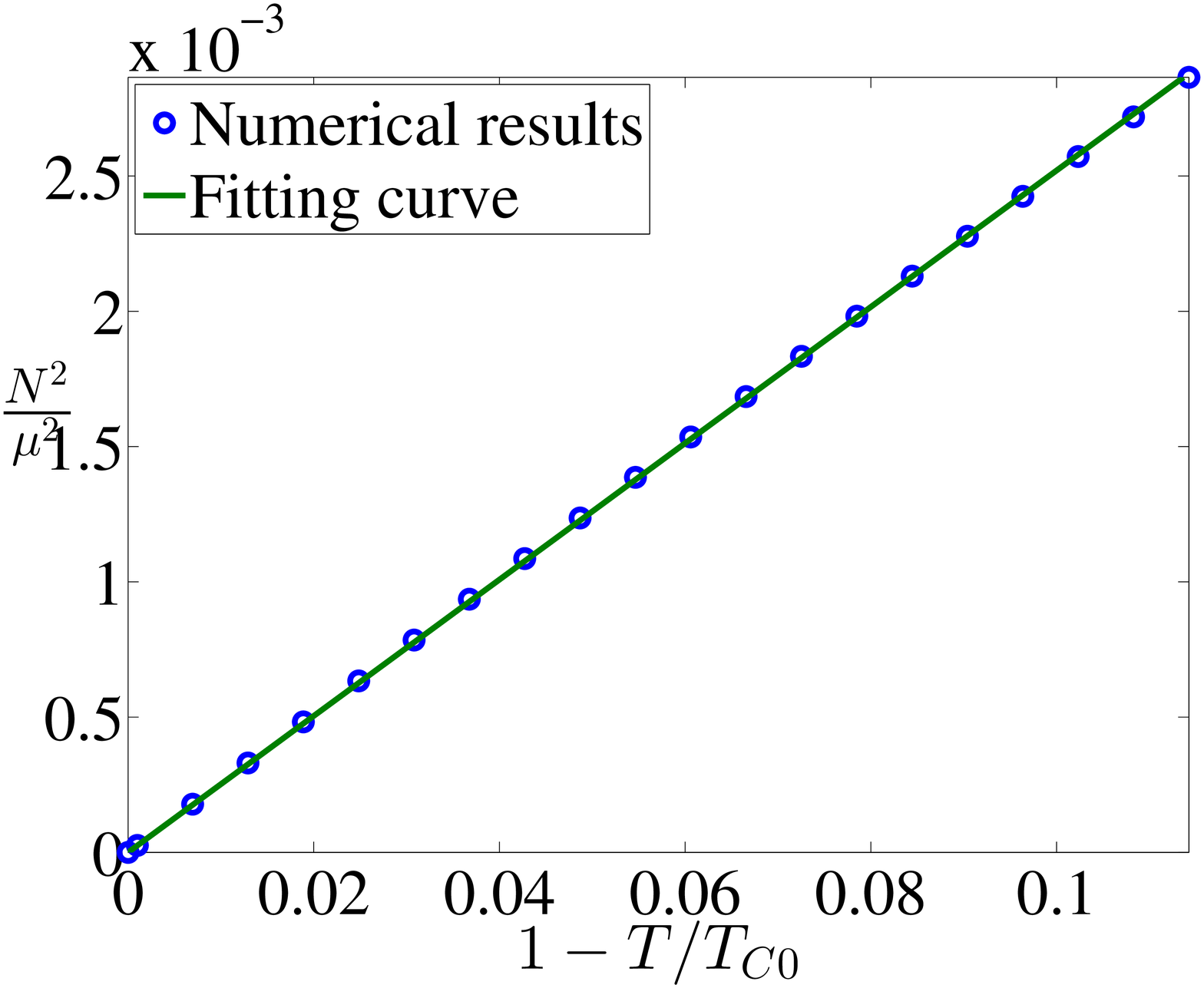}
\includegraphics[width=0.5\textwidth]{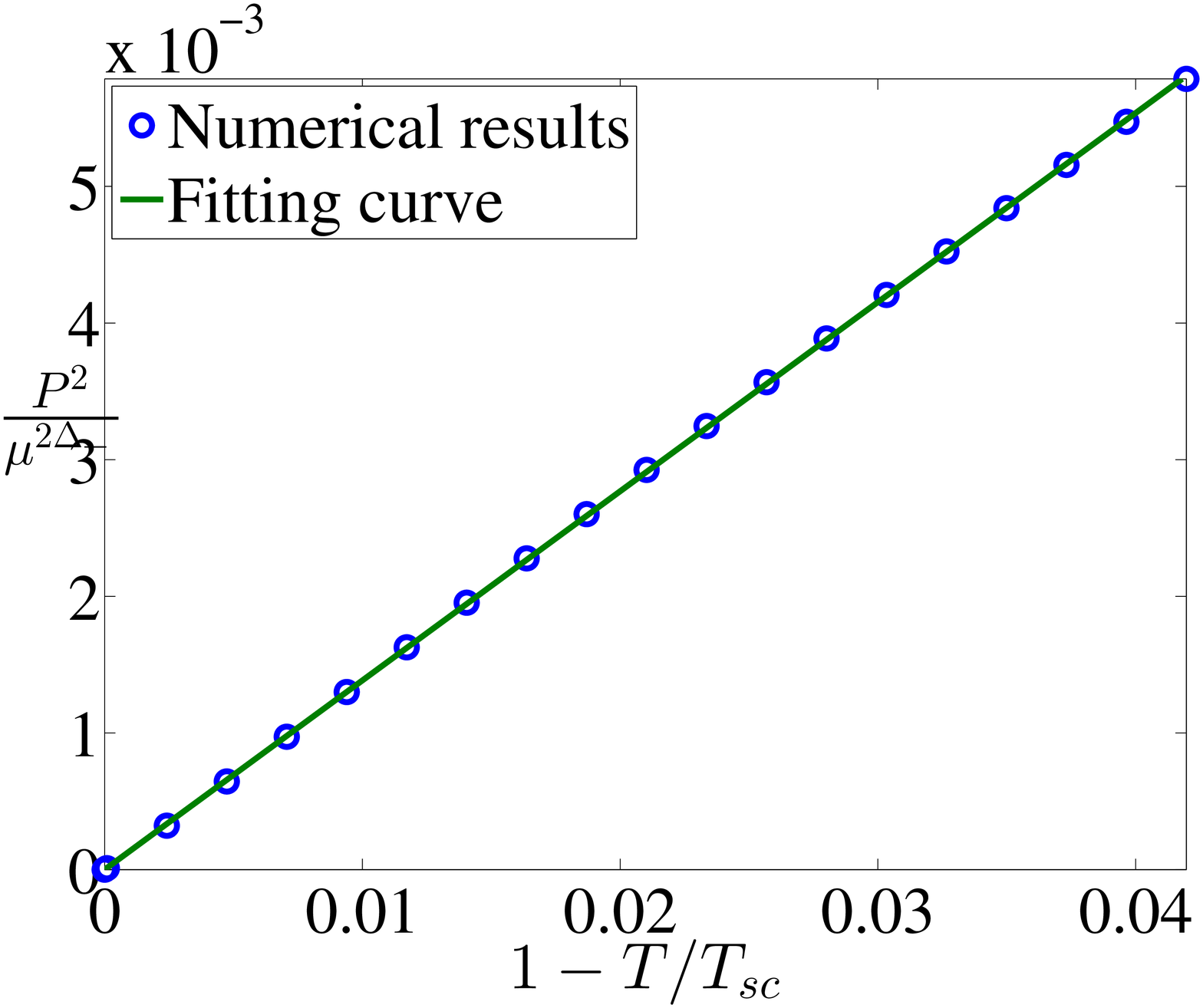}
\caption{The p-wave superconducting order parameter $P$ and spontaneous magnetic moment density $N$ with respect to temperature near the critical temperature in the Phase $D_1$.  Numerical fittings give  that $P\mu^{-\Delta_-}\simeq0.3721\sqrt{1-T/T_{sc}}$ and $N/\mu\simeq0.1588\sqrt{1-T/T_{C0}}$. Here $m_1^2=-3/16, m_2^2=-3,J=-1, q=1.3$ and $\alpha=0.1$. $\Delta_-=1+(1+\delta_1)/2$. }
\label{TG1b}
\end{figure}

In the right plot of figure~\ref{TG1}, we show the superconducting order parameter $P$ and spontaneous magnetic moment density $N$ as functions of temperature $T$ in the case of $\Theta=\pm1$ in the physical favored phase. We see that there is a second order phase transition at $T=T_{C0}$, below which the system enters  into a ferromagnetic phase. When temperature decreases below $T_{sc}$, the situations depend on the sign of $\Theta$. In the case of $\Theta>0$, there is a second order phase transitions at $T=T_{sc}$ such that the  system will transit into the p-wave superconducting ferromagnetic phase. From the curve of $N$ in the right plot of figure~\ref{TG1}, one can see that the condensation of p-wave order will increase the magnetic moment. This implies that the p-wave pair carries a nonzero magnetic moment, which contributes to the total magnetic moment of the system. In figure \ref{TG1b} we plot the behaviors of the magnetic moment and p-wave order parameter with respect to temperature in the phase $D_1$ near the critical temperature, which clearly shows a square root behavior for both quantities.

Let  us  make a brief summary for this subsection. In the case of $T_{C0}>T_{sc0}$, i.e.,  the case with the ferromagnetic phase  appearing  first with decreasing the temperature, whether the p-wave superconductivity can appear depends on the sign of $\Theta$. If $\Theta>0$, there is a critical temperature $T_{sc}$ which is lower than $T_{C0}$ but higher than $T_{sc0}$. When $T<T_{sc}$, the p-wave superconductivity can appear and the system will show the ferromagnetism and superconductivity both. Even in the near zero temperature limit, they can coexist. On the other hand, if $\Theta<0$, the p-wave superconductivity can not appear and the system will only be in a pure ferromagnetic phase. These results are summarized in  table~\ref{Tab1}.


\subsection{Ferromagnetic superconductor}

Let us now consider the other case that $T_{sc0}>T_{C0}$, i.e., the case where the p-wave superconducting phase appears first. When $T_{C0}<T<T_{sc0}$, according to the equations~\eqref{eqcomp1}, there may exist three kinds of p-wave superconducting phase. One is just the usual p-wave superconducting phase C($\{h=\rho_y=0, \rho_x\neq0\}$), the other two are new superconducting phase $E_1$ with $\{h<0,\rho_x=\rho_y\neq0\}$ and $E_2$  with $\{h<0,\rho_x=-\rho_y\neq0\}$. Thought phase $E_1$ and $E_2$ have nonzero magnetic moment, they are different with phases $D_1$ and $D_2$, because the appearance of nonzero $h$ in the former two is induced by the p-wave pair but in the latter two is spontaneously produced.  Thus we have two questions as follows. Can these three solutions exist ?  And which one  is physical favored when temperature is in the region $T_{C0}<T<T_{sc0}$?

\begin{figure}
\includegraphics[width=0.5\textwidth]{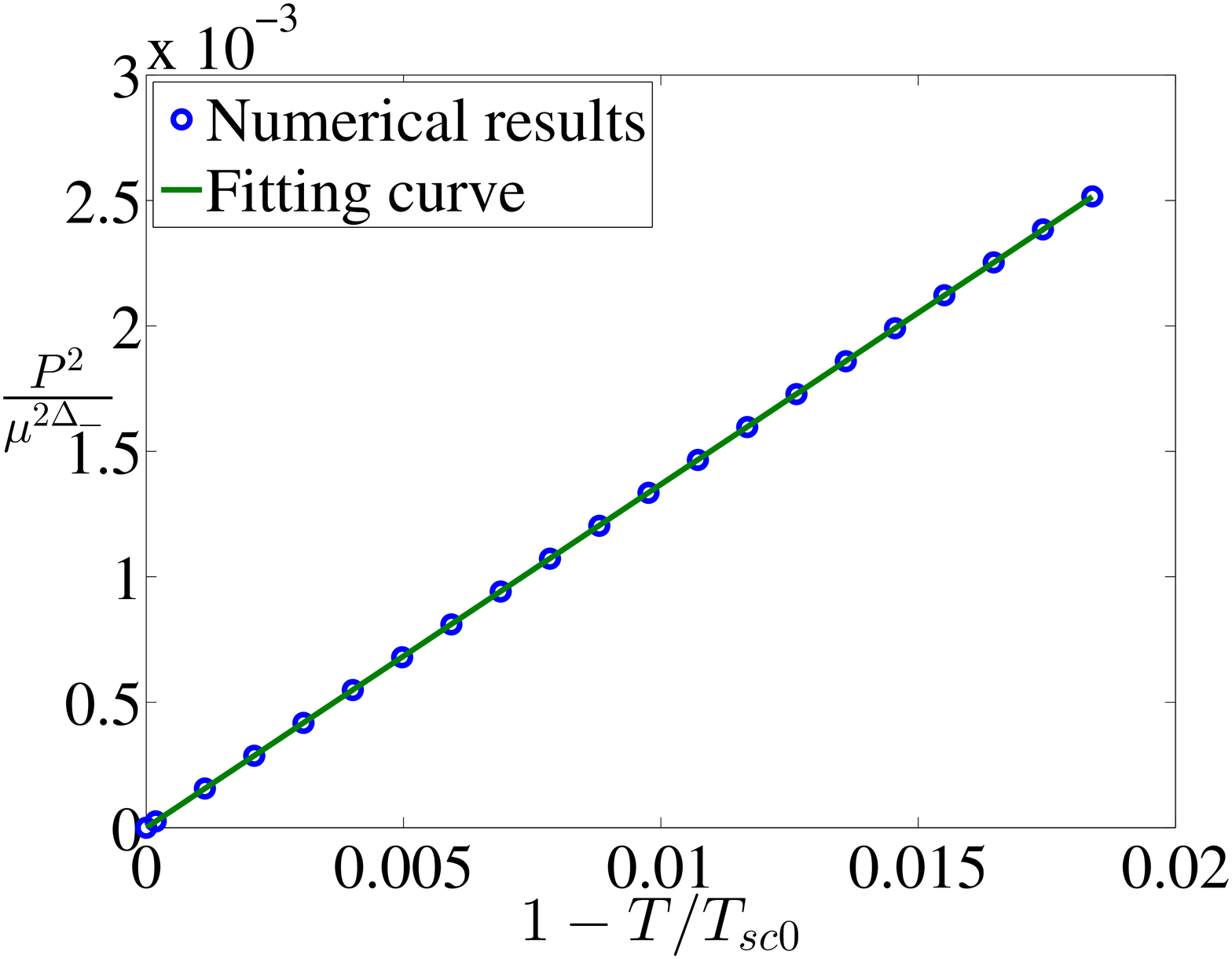}
\includegraphics[width=0.5\textwidth]{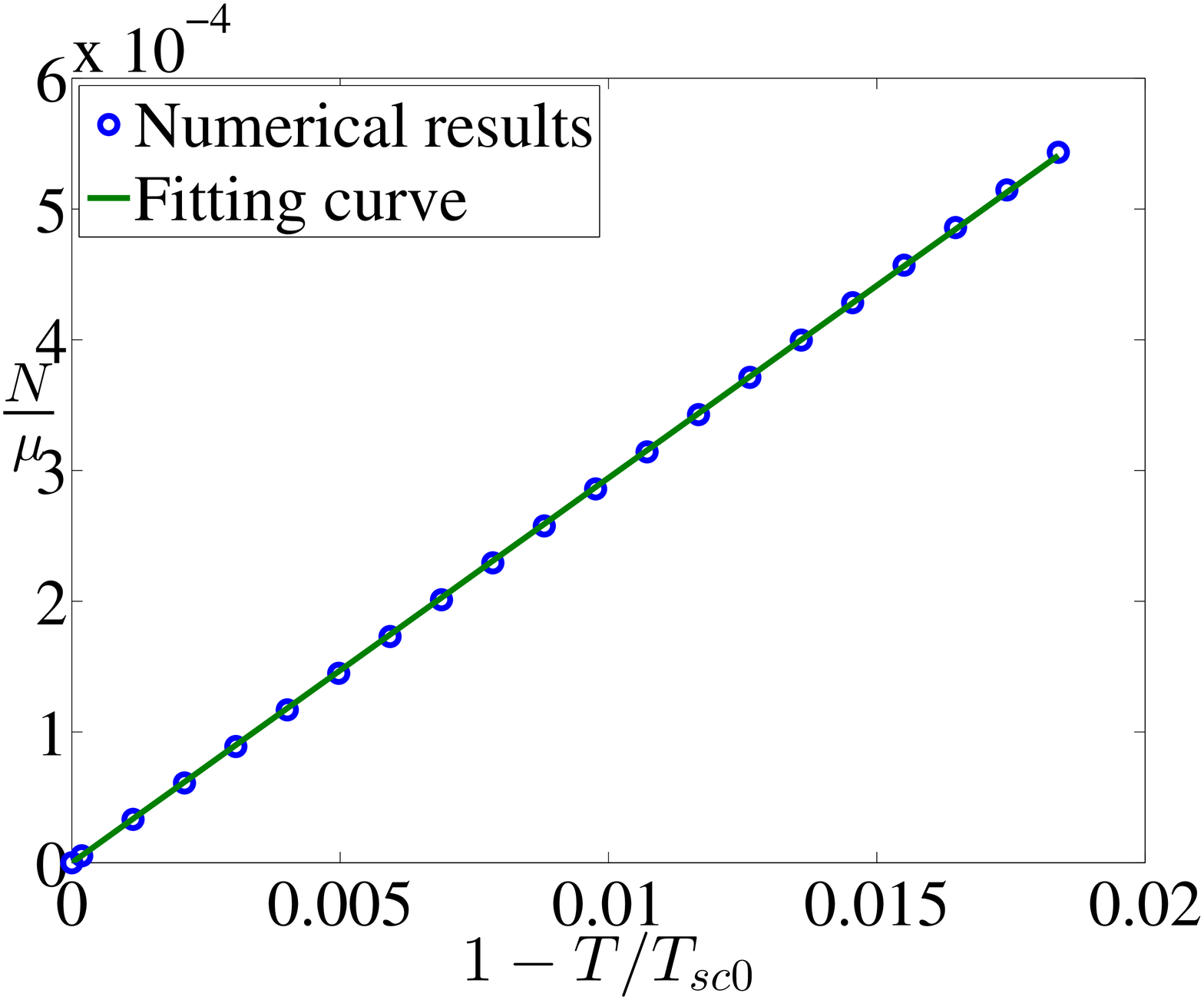}
\caption{The behaviors of  $N$ and  $P$ and near the critical temperature $T_{sc0}$ in the phase $E_1$.  Numerical fittings show that $P\mu^{-\Delta_-}\simeq0.3699\sqrt{1-T/T_{sc0}}$ and $N/\mu\simeq0.0294(1-T/T_{sc0})$. Here $m_1^2=-3/16, m_2^2=-3,J=-1, q=1.4,\Theta=1$ and $\alpha=0.1$. $\Delta_-=1+(1+\delta_1)/2$. }
\label{TG2b}
\end{figure}

Our numerical results show that the answers depend on the sign of $\Theta$. If $\Theta>0$, the phases $A$, $C$ and $E_1$ can exist and the phase $E_1$ is physical favored which has the lowest free energy. Therefore the system will show magnetism once it goes into the  p-wave superconducting phase. This case is very similar to the Anderson-Brinkman-Morel (ABM) phase in $^3$He superfluid~\cite{PWA}, where the superfluid phase is also of magnetism. Though the magnetism and superconductivity appear together, it has an essential difference from phases $D_1$ and $D_2$ just as we mentioned before. So the phase $E_1$ (and $E_2$) should be called ``magnetic superconducting'' phase rather than ferromagnetic superconducting phase. This difference can also be shown in the magnetic moment density near the phase transition point. In the phases $D_1$ and $D_2$, we see that  the $N$ shows a square root behavior with respect to $1-T/T_{C0}$ (see figure~\ref{TG1b}), while in  phase $E_1$, we see  that  $N$ has a linear relationship with respect to $1-T/T_{sc0}$ (see figure~\ref{TG2b}). On the other hand,  if $\Theta<0$, we find that the phase $E_1$ and $E_2$ do not appear and the equations~\eqref{eqcomp1} only have the trivial solution and the pure p-wave superconductivity solution. As a result, in this case, the system is in a pure p-wave superconducting phase without magnetism in the region of $T_{C0}<T<T_{sc0}$.

\begin{figure}
\includegraphics[width=0.5\textwidth]{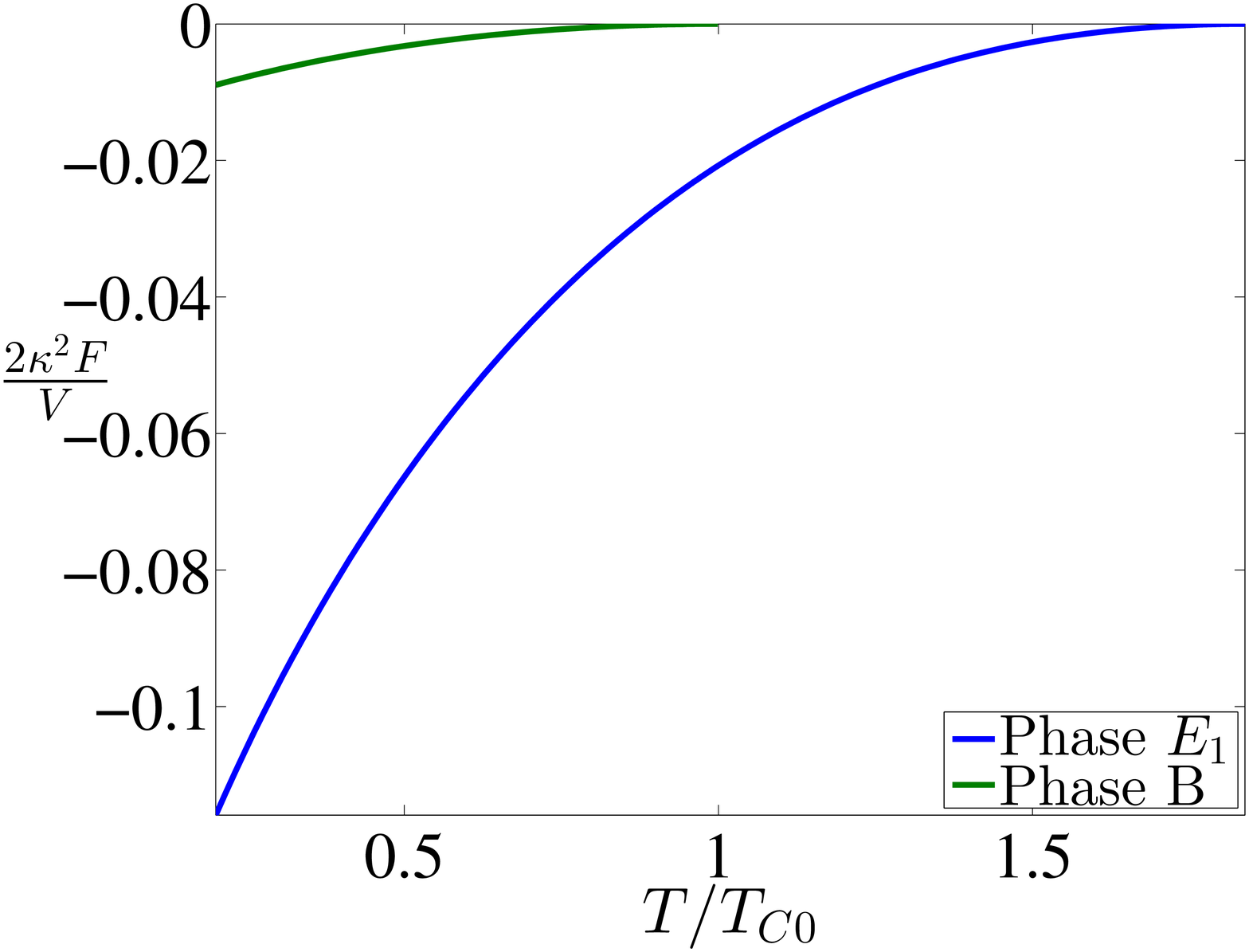}
\includegraphics[width=0.5\textwidth]{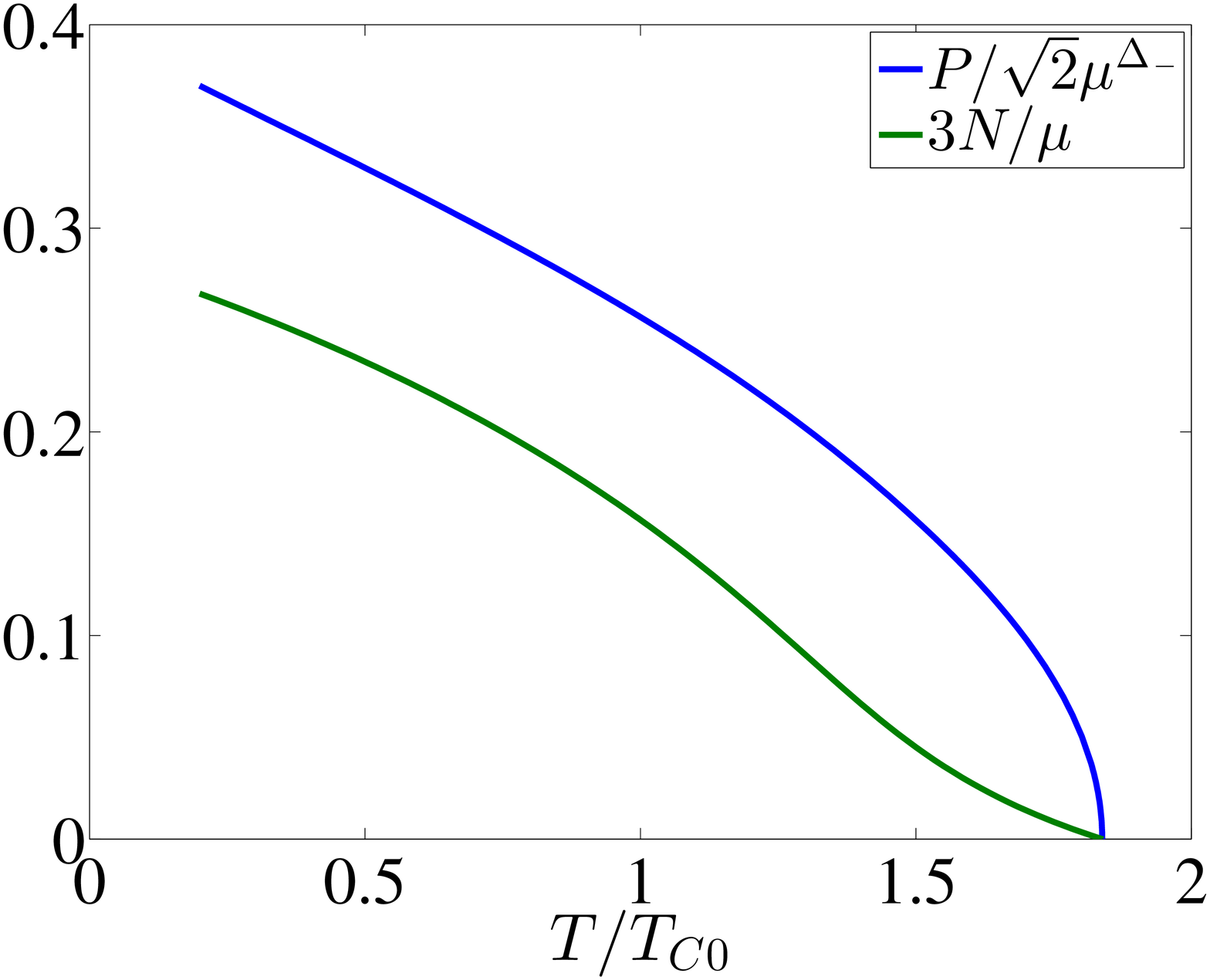}
\caption{{\bf Left:} The free energy density with respect to temperature in  phase $E_1$ and phase $B$.  {\bf Right:} The p-wave superconducting order parameter $P$ and induced magnetic moment density $N$ with respect to temperature in  phase $E_1$. Here $m_1^2=-3/16, m_2^2=-3,J=-1, q=1.4,\Theta=1$ and $\alpha=0.1$. $\Delta_-=1+(1+\delta_1)/2$. }
\label{TG2}
\end{figure}

As an example, let us consider the parameters as $m_1^2=-3/16, m_2^2=-3,J=-1, q=1.4$ (again, the results are qualitatively similar for other parameter values, the only requirement is to have
$T_{sc0} > T_{C0}$). In this case, we have  $T_{sc0}\simeq1.8383T_{C0}$. The free energy density, the p-wave superconducting order parameter $P$ and induced magnetic moment density $N$ with respect to temperature in the phase $E_1$ are shown in  figure~\ref{TG2}. Because the on-shell free energy for phases A and C is zero in the probe approximation,  we  see that phase $E_1$ has the lowest free energy. Furthermore from  figure~\ref{TG2b}, we see that the p-wave superconducting order parameter $P$ have a square root behavior with respect to $1-T/T_{sc0}$ but the induced magnetic moment density has a linear relationship with $1-T/T_{sc0}$  when $T\rightarrow T^{-}_{sc0}$.

\begin{figure}
\includegraphics[width=0.5\textwidth]{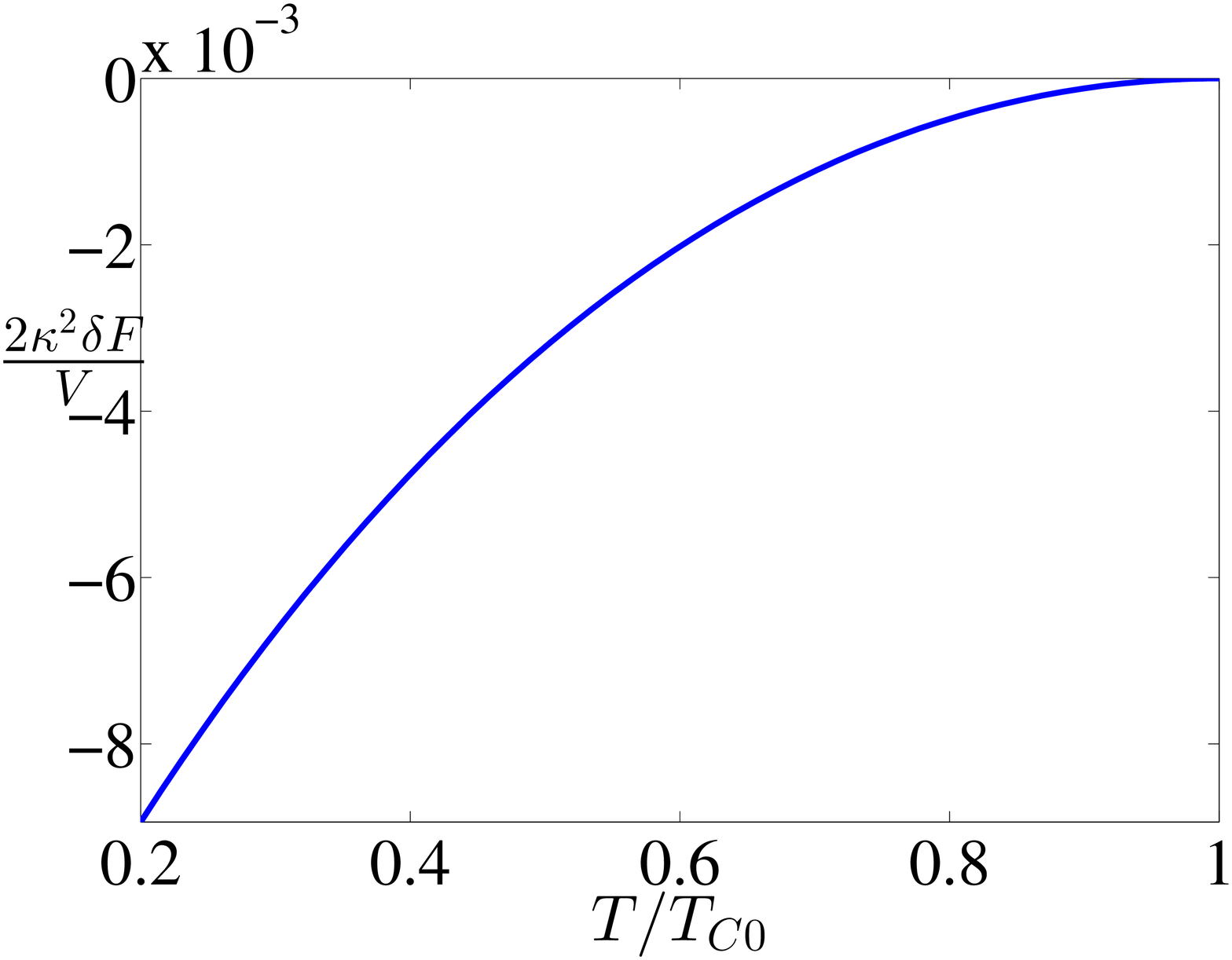}
\includegraphics[width=0.5\textwidth]{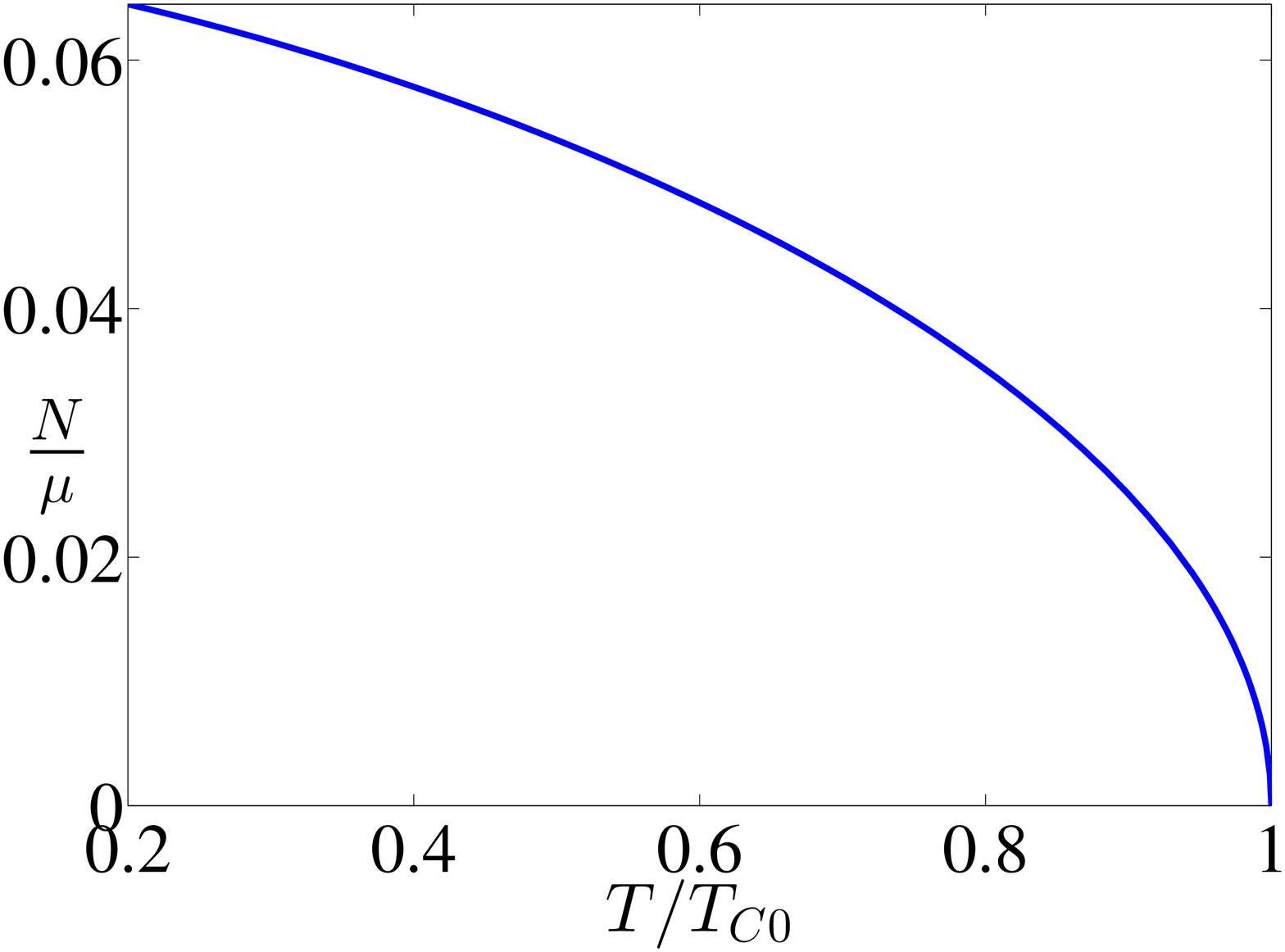}
\caption{{\bf Left:} The difference of free energy between phase C and phase B. Here $\delta F=F_{\text{phase B}}-F_{\text{phase C}}$. {\bf Right:} The magnetic moment density $N$ with respect to temperature in  phase $B$. Here $m_1^2=-3/16, m_2^2=-3,J=-1, q=1.4, \Theta=-1$ and $\alpha=0.1$. }
\label{TGN}
\end{figure}

When temperature is lower than $T_{C0}$ and $\Theta>0$, phases B can also appear. The numerical results show that the phase $E_1$ is still the lowest free energy phase (see figure~\ref{TG2} as an example). So there is no phase transition at $T=T_{C0}$. If $\Theta<0$, we have known that when $T>T_{C0}$, this system is in  a pure p-wave superconducting phase C with $\{h=\rho_y=0,\rho_x\neq0\}$. At $T=T_{C0}$, because the equation for $h$ is a homogeneous one ($\rho_y=c\rho_x$, in phase C we have $c=0$), there is critical point for $h$. However,  the equation for $c(r)$ in~\eqref{eqcomp1} restricts $c(r)$ to be 0 or $\pm1$. When $h\neq0$,  the only solution for  $c(r)$ is $c(r)=\pm1$. This indicates that in this case  the solutions for~\eqref{eqcomp1} are  either just phase $E_1$ (or $E_2$) or phase B, i.e.,  phases $\{h<0,\rho_x=\pm\rho_y\neq0\}$ or $\{h<0,\rho_x=\rho_y=0\}$. However, our numerical calculations show that such solutions $\{h<0,\rho_x=\pm\rho_y\neq0\}$ do not  exist. Thus the possible phases are only phase A, phase B and phase C. In this model, the on shell free energy is zero for phase A and C but is negative for phase B. So the phase $\{h<0,\rho_x=\rho_y=0\}$ is physical favored in the case of $\Theta<0$ when $T<T_{C0}$. The ferromagnetism can still appear from the p-wave superconducting phase at the same critical temperature $T_{C0}$, but the p-wave superconducting phase will disappear. In other words, the superconductivity and ferromagnetism can not  coexist in the case of $\Theta<0$. In  figure~\ref{TGN}, we plot the difference of free energy between phase C and phase B and magnetic moment density $N$ with respect to temperature in  phase $B$, where $\delta F=F_{\text{phase B}}-F_{\text{phase C}}$, from which we can see that the pure ferromagnetic phase has lower free energy than the pure p-wave superconducting phase.
\begin{table}
  \centering
  \begin{tabular}{|c|c|c|c|}
  \multicolumn{4}{c}{Phases in the case of $\Theta>0$}\\
    \hline
    Temperature & $T>T_{sc0}$& $T_{C0}<T<T_{sc0}$&$T<T_{C0}$\\
    \hline
  Possible& A &  A, $E_1$, C &  A, $E_1$, C, B\\
    \hline
    Physical &A& \multicolumn{2}{c|}{$E_1$}\\
    \hline
    \multicolumn{4}{c}{ }\\
  \end{tabular}
  \\

\centering
\begin{tabular}{|c|c|c|c|}
  \multicolumn{4}{c}{Phases in the case of $\Theta<0$}\\
    \hline
    Temperature & $T>T_{sc0}$& $T_{C0}<T<T_{sc0}$&$T<T_{C0}$\\
    \hline
  Possible& A &  A, C &  A, C, B\\
    \hline
    Physical &A& C& B\\
    \hline
  \end{tabular}
  \caption{The possible and physical phases in the case of $T_{sc0}>T_{C0}$. Phase A is $\{h=\rho_x=\rho_y=0\}$. Phase B is $\{h<0,~\rho_x=\rho_y=0\}$. Phase C is $\{h=\rho_y=0,~\rho_x\neq0\}$. Phase $E_1$ is $\{h<0,~\rho_x=\rho_y\neq0\}$. Phases $E_2$ is $\{h<0,~\rho_x=-\rho_y\neq0\}$. }\label{Tab2}
\end{table}

Let us make a brief summary for this subsection. In this case of $T_{C0}<T_{sc0}$, i.e., the case that the superconducting phase will appear first, the results depend on the sign of $\Theta$. If $\Theta>0$, the ferromagnetic phase can not appear but the magnetic p-wave superconducting phase will appear in region of $T<T_{sc0}$. If $\Theta<0$, the system will be in  a pure p-wave superconducting phase in the region $T_{C0}<T<T_{sc0}$ and a pure ferromagnetic phase when $T<T_{C0}$. All the results are summarized  in  table~\ref{Tab2}.

\section{Summary and discussions}
\label{sum}

In this paper, by combining the complex vector field model for the holographic p-wave  superconductor and the real antisymmetric tensor field model for the holographic ferromagnetism, we have  investigated the coexistence and competition of ferromagnetism and superconductivity in the holographic setup.  Depending on model parameters, we  found that the model shows rich phases in  low temperatures. The study is done in the probe limit, the background geometry is taken to be an AdS RN black hole with a planar horizon.

In the case of $T_{C0}>T_{sc0}$, i.e., the case where  the ferromagnetic phase  appears first, whether the p-wave superconductivity can appear depends on the sign of $\Theta$, the  interaction strength of magnetic moment of the complex vector field. If $\Theta>0$, there is a critical temperature $T_{sc}$ which is lower than $T_{C0}$ but higher than $T_{sc0}$. When temperature is higher than $T_{sc}$, the system only shows the ferromagnetism. When $T<T_{sc}$, the p-wave superconductivity can appear and the system will show ferromagnetism and superconductivity both. Because of the the spontaneous magnetization, the critical temperature of p-wave condensation is higher than the critical temperature without the ferromagnetic phase, and  increases with the increasing of interaction strength between complex vector field and antisymmetric tensor field. Even in the near zero temperature limit, the magnetism and superconductivity can coexist. But if $\Theta<0$, the p-wave superconducting state can still exist but it  is not  the lowest free energy state. So  the superconductivity  can not  appear and the system will only be in a pure ferromagnetic state.

In the case of $T_{C0}<T_{sc0}$, i.e., the case where  the superconducting phase appears first, the results also depend on the sign of $\Theta$. If $\Theta>0$, in the region of $T_{C0}<T<T_{sc0}$, the system will show the p-wave superconductivity and a kind of induced magnetism. The superconductivity and magnetism appear both, however, it is a magnetic superconducting phase rather than a ferromagnetic superconducting phase, because  the magnetic moment is not spontaneously produced.  The magnetic moment is proportional to $T_{sc0}-T$ rather than $\sqrt{T_{sc0}-T}$ near the critical
temperature. When temperature is lower than $T_{C0}$, the ferromagnetic phase B can exist, but it has higher free energy than phase $E_1$. So in the whole region of $T<T_{sc0}$, the physical favored phase is magnetic p-wave superconducting phase $E_1$.   On the other hand, if $\Theta<0$, when temperature is less than $T_{sc0}$, the system will be in the pure p-wave superconducting phase without magnetism. If temperature is lower than $T_{C0}$, the system will transit into the pure ferromagnetic phase from the pure p-wave superconducting phase. Therefore the ferromagnetism and superconductivity can not coexist in the case of $\Theta<0$.

Now  let us  discuss  some implications of our results. We have seen that the sign of $\Theta$ plays a crucial role in this model. This phenomenological parameter in~\eqref{Vrho} describes the self-interaction between the magnetic moments of complex vector field. A positive $\Theta$ means that the magnetic moments with same direction are attractive, while a negative $\Theta$ indicates  that the magnetic moments with same direction are repulsive. From  tables~\ref{Tab1} and~\ref{Tab2}, we can see that the ground state in the near zero temperature limit only depends on the sign of $\Theta$. If we translate these attraction and repulsion into the boundary theory, then our results can be understood well. Since p-wave pair is spin triplet, it can be in the state of spin-up or spin-down. So every p-wave pair carries magnetic moment of $\pm2\mu_B$. If $\Theta>0$, which means that the p-wave pair will attract the pair which has the same magnetic moment direction and repulse the one which has opposite magnetic moment. So under the influence of spontaneous magnetization, the p-wave pair will be enhanced and survive. In addition, the magnetic moment of p-wave pair will tend to align along the direction of spontaneous magnetization, which increases the total magnetic moment of the system. As a result we indeed  see  the ground state is the phase where the p-wave superconductivity and ferromagnetism coexist. However, if $\Theta<0$, the p-wave pair will repulse the pair which has the same magnetic moment direction. So in the region where superconductivity dominates, the p-wave pair will align without net magnetism and the system is  in a pure p-wave superconducting phase. When $T<T_{C0}$, the ferromagnetism will appear. Under the influence of spontaneous magnetization, the magnetic moment of p-wave  pair will be compelled to align the same or opposite (depends on the value of $\alpha$) direction of spontaneous magnetization, which leads to the magnetic moment of p-wave pair has same direction. But the p-wave pairs which have same magnetic moment directions will repulse each other, so the p-wave pair is not stable and will be de-paired. Thus the system can  only be in the ferromagnetic phase.

Finally let us make some additional comments on superconducting ferromagnetic materials, since the Curie temperature is higher than superconducting critical temperature in general. Although superconductivity in ferromagnets was predicted more than 30 years ago, it took many years before the first material UGe$_2$ was discovered and the research in superconducting ferromagnets has just begun recently. The main reasons why we say superconducting state in superconducting ferromagnets is unconventional are (i) Cooper pairing carry magnetism and (ii) the gap structure of superconducting  has a lower symmetry than the crystal lattices~\cite{A.de}. Let G represent the point-group symmetry of the lattice, T denote time reversal symmetry, and U(1) be the gauge symmetry. In the paramagnetic state ($T>T_{C0}>T_{sc}$) the symmetry group is given by $G\times T\times U(1)$. In the ferromagnetic phase ($T<T_{C0}$) time-reversal symmetry is broken, and in the superconducting phase ($T<T_{sc}<T_{C0}$) gauge symmetry is broken as well. The coexistence of such two critical phenomena offers an attractive playground for the investigation of new phenomena, like the elusive spontaneous vortex lattice, the influence of spin-triplet superconductivity on the ferromagnetic domain size, control of tunneling currents by magnetization and so on. Also, it is  a central issue in the understanding of superconductivity itself by the interplay of magnetism and superconductivity. Research on superconducting ferromagnetic materials will help us to expound how magnetic fluctuations can arouse superconductivity. This fundamental insight might turn out to be crucial in  designing new superconducting materials with high transition temperatures. We hope that the holographic model in this paper or the correspondence of AdS/CMT can give some helpful guidance in future.

\section*{Acknowledgements}
This work was supported in part by the National Natural Science Foundation of China  with grants No.11035008, No.11375247 and No.11435006.

\appendix
\section{The method to compute $\alpha_c$}
\label{app1}
In this appendix, we will give the method to compute the value of $\alpha_c$. We need to solve following equations in the zero temperature case
\begin{equation}\label{eqT01}
\begin{split}
h''+\frac{f'}fh'+\left(\frac{Jh^2}{r^6f}-\frac{2f'}{rf}-\frac4{r^2}-\frac{m_2^2}{fr^2}\right)h=0,\\
\rho_x''+(\frac{f'}f+\frac2r)\rho_x'+\left(\frac{q^2\phi^2}{r^4f^2}-\frac{m_1^2}{fr^2}+\frac{h\alpha}{fr^4}\right)\rho_x=0.
\end{split}
\end{equation}
Here
\begin{equation}\label{fphi}
f(r)=1-4/r^3+3/r^4,~~~\phi(r)=\sqrt{3}(1-1/r).
\end{equation}
By analyzing the behavior of $h$ and $\rho_x(r)$ near the horizon, we find the regular form for them are,
\begin{equation}\label{regT0}
h(r)=h_0+h_1(r-1)^{\beta_1}+\cdots,~~~\rho_x=(r-1)^{\beta_2}[1+\rho_{x1}(r-1)+\cdots].
\end{equation}
Substituting them into  equations~\eqref{eqT01}, the leading orders of the solutions give  the coefficients in~\eqref{regT0} as
\begin{equation}\label{coeff1}
\begin{split}
h_0=-\sqrt{m_2^2/J},~~~~\beta_1=-\frac12+\frac16\sqrt{9-12m_2^2}\\
\beta_2=-\frac12+\frac16\sqrt{9-6\alpha h_0+6m_1^2-3q^2},~~\rho_{x1}=-\frac{\beta_2}{\beta_2+1}.
\end{split}
\end{equation}
Taking~\eqref{coeff1} and \eqref{regT0} , we can integrate equations \eqref{eqT01}  from the horizon to the AdS boundary and  get the solutions outside the horizon. When the values of $m_1^2,~m_2^2,~q$ and $J$ are given, we can treat $h_1$ and $\alpha$ as shooting parameters to match the boundary conditions $h_+=\rho_{x+}=0$. Since the equation for $h$ is dependent of  $\rho_x$, we can solve $h$ first. Numerical result shows that $h_1\simeq15.40714197$. Then we put the solution $h$ into the equation of $\rho_x$ and treat  $\alpha$ as a shooting  parameter to match the boundary condition~$\rho_{x+}=0$. Finally  we find that there is solution with $\alpha_c\simeq0.57649$.


\end{document}